\documentclass[aps,pre,showpacs,twocolumn,superscriptaddress]{revtex4}

\usepackage{amsmath,amssymb} \usepackage{graphicx}
\usepackage{psfrag,color}

 \newcommand{\mbf}{\mathbf}
\newcommand{\deriv}[2]{\frac{\partial #1}{\partial #2}}
\renewcommand\mark[1]{\bgroup\color{red}\bfseries{[#1]}\egroup}

\newcommand{\D}{\displaystyle}

\begin{document}

\title{Statics and Dynamics of the Wormlike Bundle Model}

\author{Claus Heussinger}\affiliation{Laboratoire de Physique de la Mati\`ere
  Condens\'ee et Nanostructures Universit\'e Lyon 1, CNRS, UMR 5586 Domaine
  Scientifique de la Doua F-69622 Villeurbanne Cedex, France}\affiliation{Arnold
  Sommerfeld Center for Theoretical Physics and CeNS, Department of Physics,
  Ludwig-Maximilians-Universit\"at M\"unchen, Theresienstrasse 37, D-80333
  M\"unchen, Germany}

\author{Felix Sch\"uller}\author{Erwin Frey}\affiliation{Arnold Sommerfeld Center for Theoretical Physics and CeNS,
Department of Physics, Ludwig-Maximilians-Universit\"at M\"unchen,
Theresienstrasse 37, D-80333 M\"unchen, Germany}

\begin{abstract}

  Bundles of filamentous polymers are primary structural components of a broad
  range of cytoskeletal structures, and their mechanical properties play key
  roles in cellular functions ranging from locomotion to mechanotransduction and
  fertilization. We give a detailed derivation of a wormlike bundle model as a
  generic description for the statics and dynamics of polymer bundles consisting
  of semiflexible polymers interconnected by crosslinking agents. The elastic
  degrees of freedom include bending as well as twist deformations of the
  filaments and shear deformation of the crosslinks.  We show that a competition
  between the elastic properties of the filaments and those of the crosslinks
  leads to renormalized effective bend and twist rigidities that become
  mode-number dependent. The strength and character of this dependence is found
  to vary with bundle architecture, such as the arrangement of filaments in the
  cross section and pretwist. We discuss two paradigmatic cases of bundle
  architecture, a uniform arrangement of filaments as found in F-actin bundles
  and a shell-like architecture as characteristic for microtubules. Each
  architecture is found to have its own universal ratio of maximal to minimal
  bending rigidity, independent of the specific type of crosslink induced
  filament coupling; our predictions are in reasonable agreement with available
  experimental data for microtubules. Moreover, we analyze the predictions of
  the wormlike bundle model for experimental observables such as the
  tangent-tangent correlation function and dynamic response and correlation
  functions. Finally, we analyze the effect of pretwist (helicity) on the
  mechanical properties of bundles. We predict that microtubules with different
  number of protofilaments should have distinct variations in their effective
  bending rigidity.

\end{abstract}

 \date{\today}

\maketitle

\section{Introduction}

Bundles of filamentous polymers like F-actin form primary structural components
of a broad range of cytoskeletal structures including stereocilia, filopodia,
microvilli, cytoskeletal stress fibers, or the sperm acrosome. Actin-binding
proteins allow the cell to tailor the dimensions and mechanical properties of
the bundles to suit specific biological functions. In particular, the mechanical
properties of these bundles play key roles in cellular functions ranging from
locomotion~ \cite{mogilnerBPJ2005,atilganBPJ2006,vignejvic2006JCellBiol} to
mechanotransduction~\cite{hudspeth1977PNAS} and fertilization~
\cite{schmid2004Nature}. In view of this ubiquity, a detailed understanding of
bundle mechanics is fundamental to gaining a mechanistic understanding of
cellular function~\cite{howard2008review}. Quantifying the governing mechanical
principles of these fundamental cytoskeletal constituents could also prove
valuable in the design of biomimetic nanomaterials.

In-vitro experiments recently investigated the role of actin-binding proteins
like fascin, $\alpha$-actinin and $I$-plastin in mediating bundle
mechanical properties~\cite{Claessens2006}.  Already an inspection of bundle
conformations from fluorescence microscope images makes it evident that the
properties of the various crosslinking proteins must be quite distinct. While
bundles formed by fascin show a compact form and remain straight over several
microns, bundles formed by $\alpha$-actinin or filamin are wiggly and very lose~
\cite{pelletier03,lielegWLB2007,schmollerPRL2008}. The mechanical properties of
actin bundles formed by different crosslinking proteins was quantified by a
fluctuation analysis~\cite{Claessens2006}, which measures the magnitude of their
thermal fluctuations. It was found that the apparent bundle bending stiffness
can be varied over a substantial range by changing the type and relative
concentration of the crosslinker.

\begin{figure}[ht]
 \begin{center}
   \includegraphics[width=0.9\columnwidth]{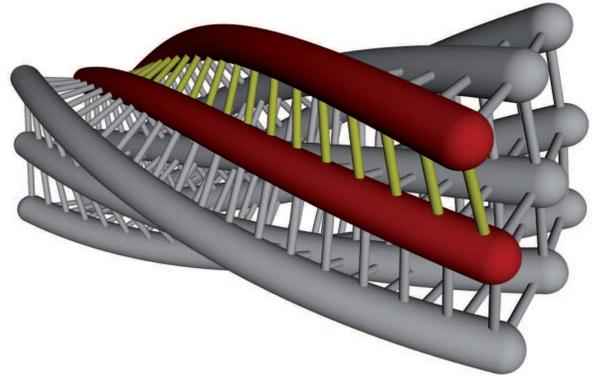}
\end{center}
\caption{\emph{Wormlike bundle model.} We consider bundles that consist of
  regular arrangements of filament. These are assumed to be locked in place by
  crosslinking proteins. When the bundle bends and twists in space, the
  filaments start to slide along each other. This effect leads to shear
  deformation in the crosslinks.}
  \label{fig:bundle_illustration}
\end{figure}

These intriguing mechanical properties can be understood in terms of
the wormlike bundle model (WLB), which describes bundles as an
assembly of semiflexible filaments interconnected by crosslinking
proteins~\cite{batheBPJ2008,heussingerWLB2007}; for an illustration of
the bundle architecture see Fig.~\ref{fig:bundle_illustration}. Unlike
the standard wormlike chain model (WLC)~\cite{kratky49,saito67}, the
wormlike bundle model (WLB) exhibits a state-dependent bending
stiffness~\cite{batheBPJ2008} that derives from a generic competition
between the bending and twist stiffness of individual filaments and
their relative motion mediated by the stiffness of the
crosslinkers. An important aspect of the WLB model is that crosslinks
may be very efficient in constraining the lateral excursions of
filaments within the bundle but much less so in inhibiting their axial
motion.  This is possible as the relative axial sliding of two
crosslinked filaments probes not only the elastic properties of the
crosslinking protein but also those of the binding domain at which the
protein is attached to the filament. The latter may be of quite
different rigidity than the protein itself.  Unfortunately there are
no single-molecule experiments yet which would quantify the mechanical
and binding properties of actin crosslinking proteins attached to a
pair of F-actin filaments~\footnote{The very rare experiments that
  have tried to measure the binding strength of an actin crosslinker
  have been attempted with surface adsorbed crosslinkers or surface
  adsorbed
  filaments~\cite{ferrer2008PNAS,leeCellBioEng2009,miyataBiochimActa1996}. However,
  to understand the contributions of actin crosslinkers in a
  dynamically strained cytoskeleton it is essential to measure the
  biophysical properties of the full, freely suspended crosslink on a
  single molecule level.}. In the WLB model the mechanical properties
of the crosslinks are described by a single shear-stiffness $k_\times$
for the relative sliding of the constituent filaments.

An important finding within the wormlike bundle model is that the  
mechanical properties of bundles can be classified into three distinct  
bending regimes that are mediated by both crosslink type and equally  
importantly by bundle dimensions, namely diameter and length~
\cite{batheBPJ2008}. Taking into account the mechanical properties of  
filaments and crosslinker at a microscopic level is a virtue of the  
model making it quite verstile. It reduces to a well defined continuum  
limit but is equally applicable to bundles with as few as two  
filaments. This microscopic perspective may prove a valuable starting  
point to address more complex questions related to bundle mechanical  
properties. This may include problems such as disorder, lattice  
defects \cite{govPRE2008}, or filament
fracture~\cite{guild2003JCellBiol}.

\begin{figure}[h]
 \begin{center}
   \includegraphics[width=0.47\columnwidth]{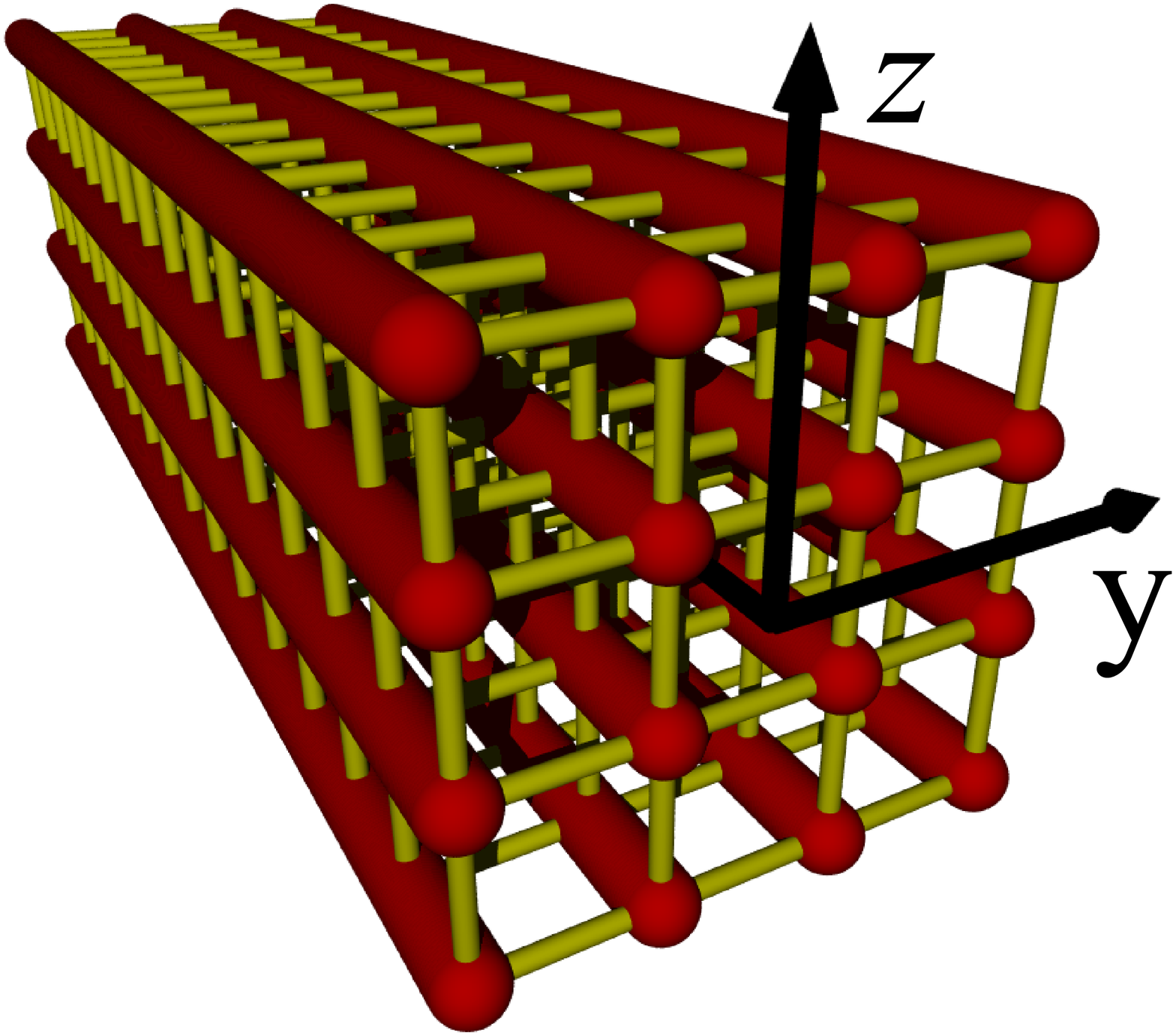}
   \hfill
   \includegraphics[width=0.47\columnwidth]{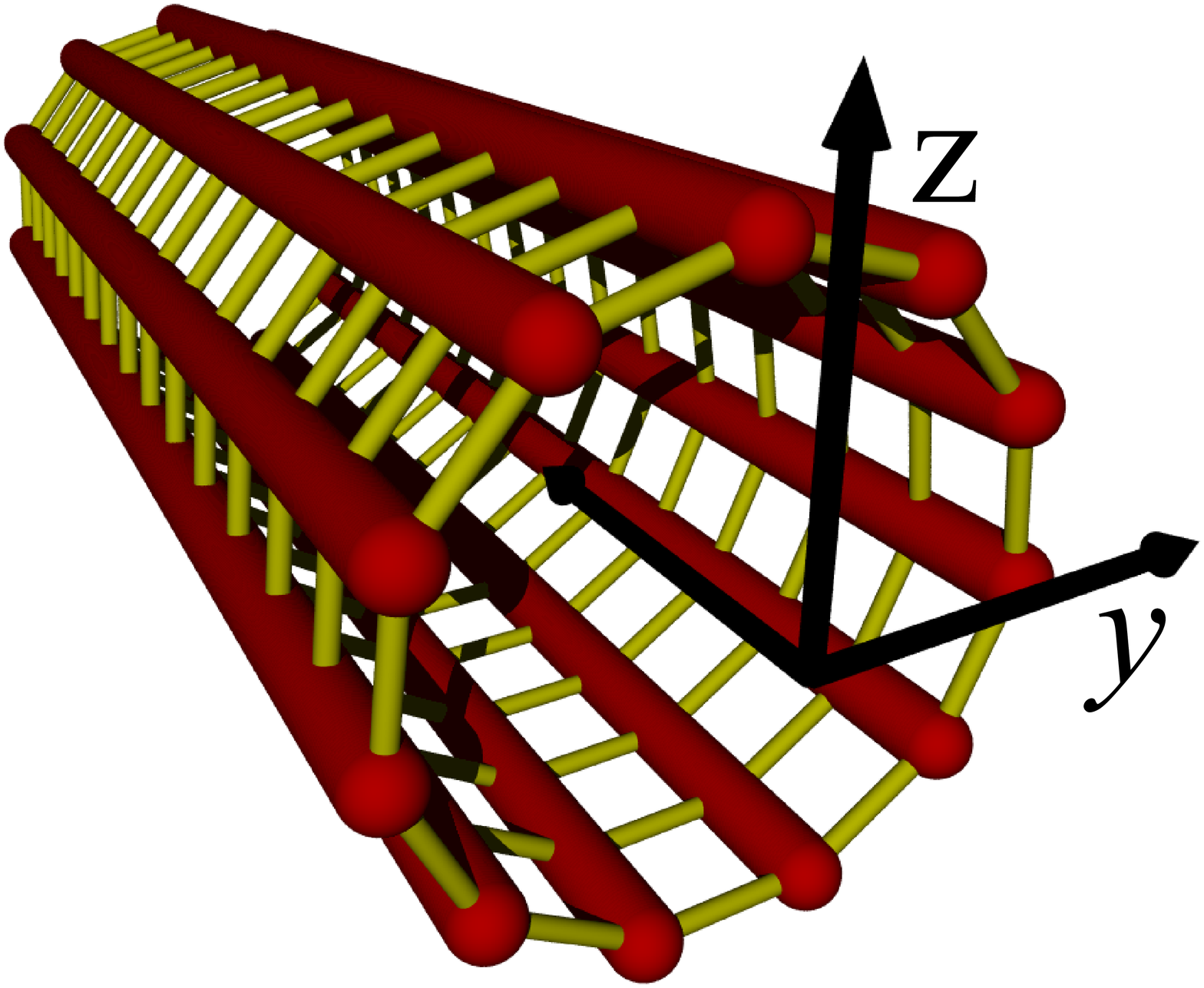}
   \includegraphics[width=0.4\columnwidth]{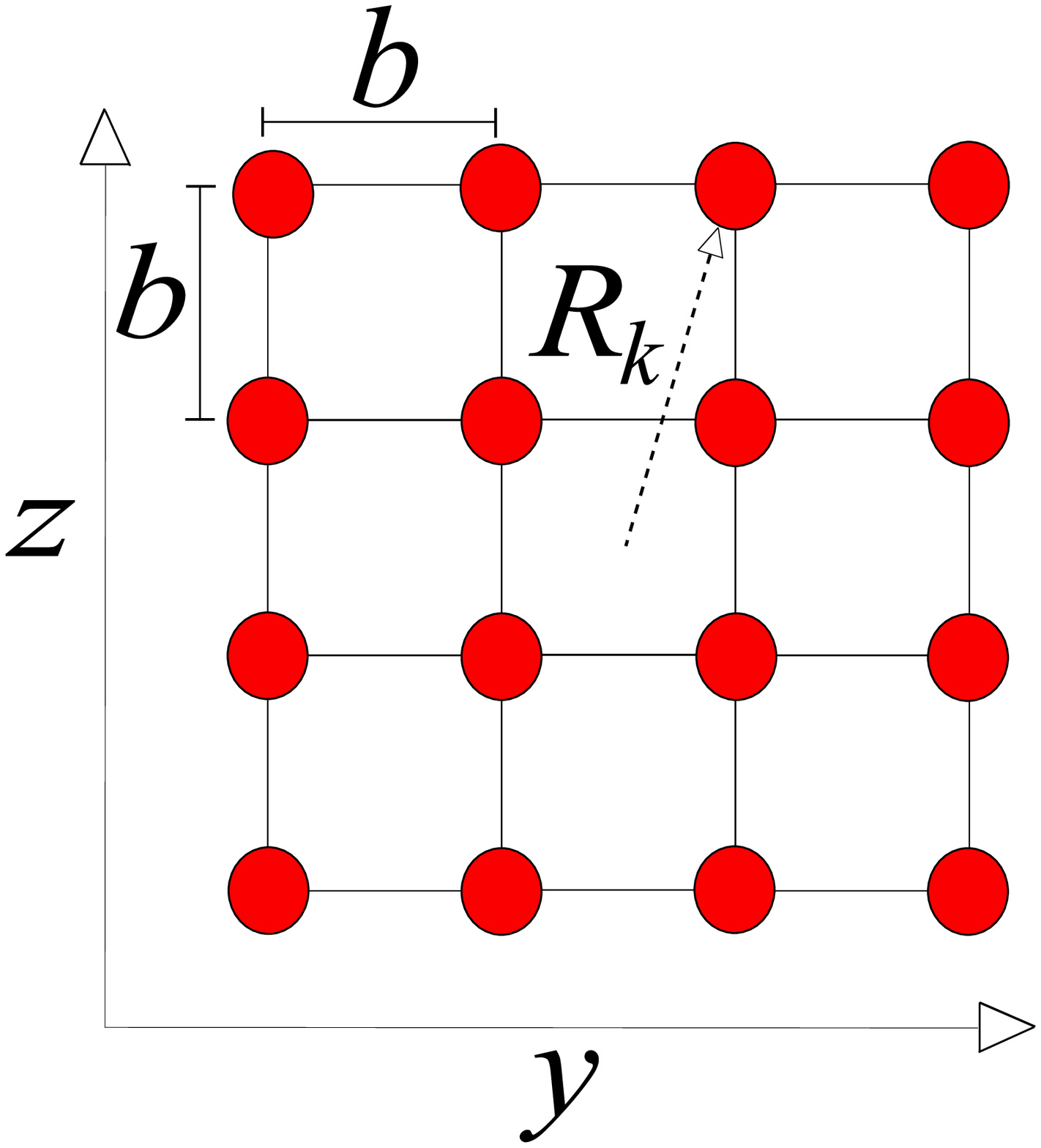}
   \hfill
   \includegraphics[width=0.53\columnwidth]{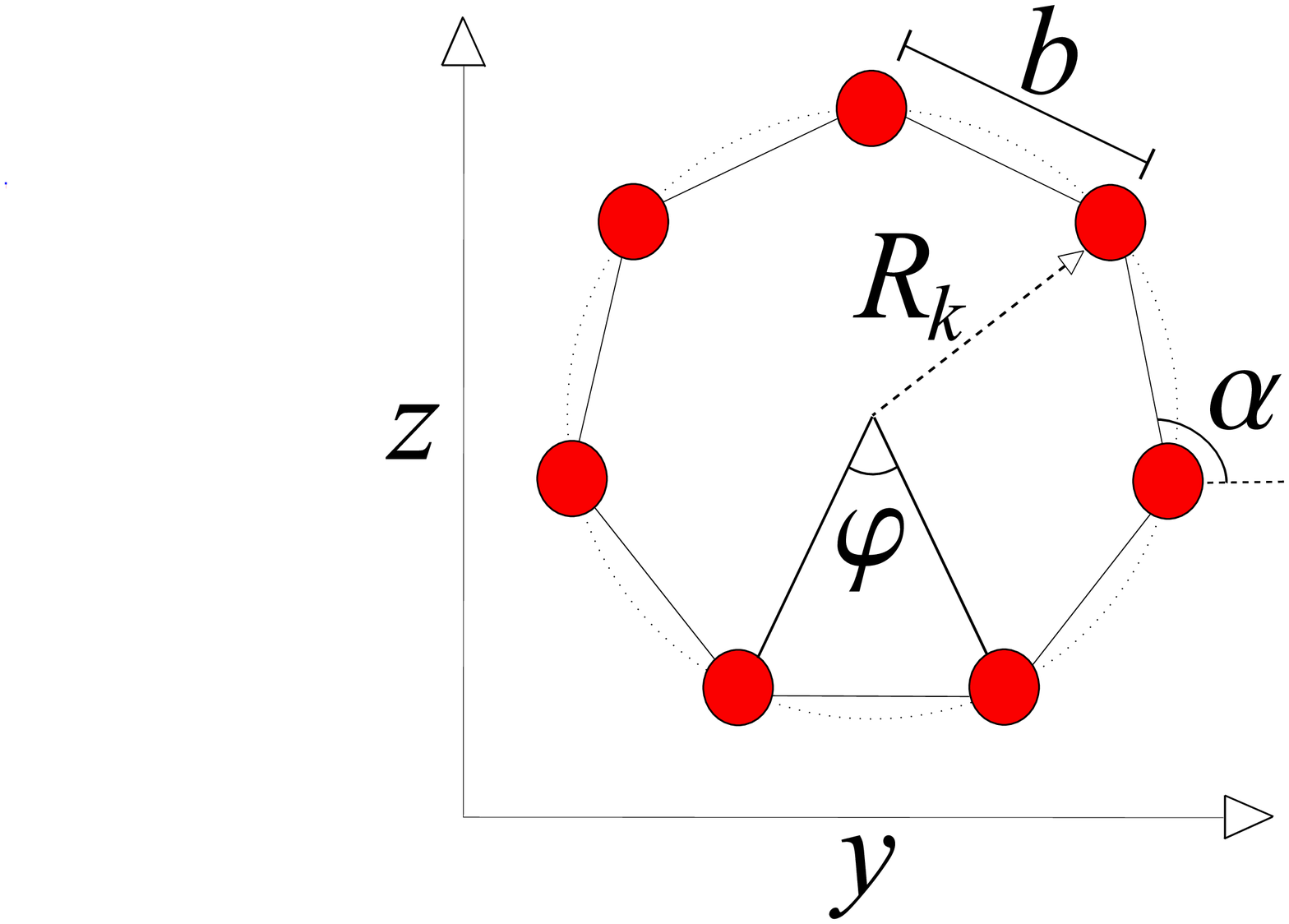}
\end{center}
\caption{llustration of the bundle geometries considered. The position of the  
$k$th filament in the cross-sectional plane ($(y,z)$-plane) is given  
by ${\bf R}_k$. \emph{F-actin bundle architecture} (left): filaments  
are arranged on a square lattice. \emph{Microtubule architecture}  
(right): filaments are arranged on the surface of a cylinder.}
  \label{fig:quantities_explained}
\end{figure}

While previously we have provided a formulation only for plane-bending
in two dimensions~\cite{heussingerWLB2007}, here we give a full
derivation of the WLB Hamiltonian in three spatial dimensions. This
includes bending as well as twist deformations. We explore the
predictions of the WLB model for experimental observables such as the
tangent-tangent correlation function and dynamic response
functions. Moreover, we discuss the effects of different bundle
geometries on their mechanical properties. In this respect, we will
view microtubules as a bundle of (proto-)filaments arranged on
the surface of a cylinder; compare
Fig.~\ref{fig:quantities_explained}. This shell-like bundle
architecture is contrasted with a uniform distribution of filaments as
found for F-action
bundles~\cite{claessensPNAS2008,purdy2007PRL}. Finally, we discuss how
helicity influences bundle mechanics.

\section{Model definition}

We consider bundles of length $L$ that consist of $N$ parallel filaments. While
the filaments may form a disordered lattice structure, we will focus our
attention here to the cases of regular arrangements. In particular we will treat
in detail the square-lattice and the cylindrical tube (see
Fig.~\ref{fig:quantities_explained}).

Each filament is modeled mechanically as an extensible worm-like polymer with
stretching stiffness $k_s$, bending rigidity $\kappa_b$ and twist rigidity
$\kappa_t$.  Filaments are irreversibly crosslinked to their nearest-neighbors
by crosslinks with mean axial spacing $\delta$. The crosslinks are modeled to be
compliant in shear along the bundle axis with finite shear stiffness $k_\times$,
and to be inextensible transverse to the bundle axis, thus constraining the
interfilament distance, $b$, to be constant (see
Fig.~\ref{fig:quantities_explained}). This assumption, which neglects crosslink
stretching deformations, is based on the recognition that the shearing mode
involves deformation of the crosslink \emph{and} its binding domain to the
filament. The resulting stiffness may indeed be much lower than that of a
crosslink in isolation.

Filament stretching is characterized by the axial displacement $u_k(s)$ of
filament $k$ at axial position $s$ along the backbone. To describe bundle
bending and twisting we define $\{\mbf{d}_1,\mbf{d}_2,\mbf{d}_3\}$ to be a
material frame fixed to the bundle central line at each arclength position $s$.
The vector $\mbf{d}_3\equiv \mbf{t}$ is the tangent to the space curve traced
out by the central line, while the two vectors $\mbf{d}_1$ and $\mbf{d}_2$ lie
within the cross-section of the bundle. The position of each filament in the
cross-section is parametrized in terms of a vector
$\mbf{R}_k(s)=A_k\mbf{d}_1(s)+B_k\mbf{d}_2(s)$, where $A_k$ and $B_k$ are the
material-frame coordinates of filament $k$; they are constants independent of
arclength $s$ and deformation of the central line (see
Fig.~\ref{fig:quantities_explained}). As one moves along the bundle backbone the
material frame rotates according to Frenet-Seret equations,
$\partial_s\mbf{d}=\mbf{\Omega}\times\mbf{d}$. The rate of rotation is given by
the generalized curvatures $\mbf{\Omega}=(\Omega_1(s),\Omega_2(s),\Omega_3(s))$,
which, in addition to the axial displacement $u_k$, represent the basic
kinematic degrees of freedom of the bundle.

\subsection{The WLB Hamiltonian}\label{sec:wlb-hamiltonian}

Neglecting all nonlinear effects we are now going to develop a simple expression
for the bundle energy which is harmonic in its degrees of freedom, axial
displacement $u_k$ and generalized curvatures $\Omega_\alpha$.

This WLB Hamiltonian consists of three contributions, $H_{\rm WLB}=H_{\rm
  0}+H_{\rm stretch}+H_{\rm shear}$. The first term corresponds to the standard
wormlike-chain Hamiltonian
\begin{equation}\label{eq:H0_Omega} H_0 = \frac{N}{2}\int_0^Lds \left[
\kappa_b\left(\Omega_1^2+\Omega_2^2\right) + \kappa_t\Omega_3^2 \right]\,.
\end{equation}
Writing this we assume that each filament follows effectively the same
space-curve as the center-line. While, in general, one should account for the
curvatures, $\mbf{\Omega}_{k}$, of each individual filament separately, this
would only lead to correction factors that can be neglected for our purposes.
Consider, for example, planar bending of the central line, $\Omega_1= 1/\rho$,
where $\rho$ is the radius of curvature. The filaments that lie at a distance
$R$ away from the central line naturally have a different radius of curvature,
$\rho\pm R$, and thus a different bending energy. The magnitude of the
correction term relative to $\Omega_1^2$ is small since it scales as
$(R/\rho)^2$, i.e. we assume the typical curvatures to be smaller than the
bundle radius (for more details see Appendix~\ref{sec:relat-betw-omeg}).

As to the twist degree of freedom, in Eq.~(\ref{eq:H0_Omega}) we do not allow
for the possibility of relative twisting of the individual filaments (see
Section~\ref{sec:pretwisted-bundles} and Appendix \ref{sec:relat-betw-omeg} for
a discussion of this effect). We assume that twist is only due to the
``bundle-twist'' $\Omega_3$ of the central line. This assumption is reasonable
in tightly bound bundles, where the filaments are connected by many crosslinks.
In this state the filaments and their relative orientation can be assumed to be
locked-in by the crosslinker.

The second term in the Hamiltonian, $H_{\rm stretch}=\sum_k H_{\rm stretch}^k$,
accounts for filament stretching. It depends on the difference in axial
displacement, $u_k$, between two crosslinks at arclength positions $s_i$ and
$s_i+\delta$, respectively.
\begin{eqnarray}\label{eq:Hstretch}
  H_{\rm stretch}^k &=& \frac{k_s}{2}\sum_{i}\left[u_k(s_i+\delta)-u_k(s_i)\right]^2\nonumber\\
  &\to&\frac{k_s\delta}{2}\int_0^Lds \left(\deriv{u_k}{s}\right)^2\,,
\end{eqnarray}
where we have performed the continuum limit $\sum_i\to\int ds/\delta$ to arrive
at the second line. The spring constant $k_s(\delta)$ is the single filament
stretching stiffness on the scale of the crosslink spacing $\delta$.

The particular form for $k_s$ depends on the system under consideration.  For
high crosslink concentrations (small $\delta$), the segment behaves as a
homogeneous elastic beam, characterized by a Youngs modulus $E$ and $k_s^{\rm
  beam}\sim Eb^2/\delta$. The combination $k_s\delta$ that enters the
Hamiltonian is independent of $\delta$, as it should: the mechanical stretching
stiffness of a beam cannot depend on the properties of the crosslinks. 

For small concentrations of crosslinks (large $\delta$) entropic effects become
relevant and the stretching stiffness is that of a thermally fluctuating
wormlike chain with persistence length $l_p$. In this case one has $k_s^{\rm
  entr}\sim \kappa_bl_p/\delta^4$, which implies that the combination
$k_s\delta\sim \delta^{-3}$ \emph{does} depend on the crosslink spacing
$\delta$. This is related to the fact that the formation of a crosslink
suppresses thermal undulations (reduces entropy) and thus increases the entropic
stretching stiffness. Equating both stretching stiffnesses, $k_s^{\rm beam}\sim
k_s^{\rm entr}$ one finds the critical crosslink concentration, $\delta_c^3\sim
b^2l_p$, at which the cross-over from enthalpic to entropic elasticity takes
place.

In the case of microtubules, which will be treated in
Section~\ref{sec:micr-nanot}, the crosslink spacing $\delta$ is given by the
tubulin-size and the stretching stiffness is modeled as for an elastic beam.

The third energy contribution, $H_{\rm shear}=\sum_{lk} H_{\rm shear}^{lk}$,
results from the crosslink-induced coupling of neighboring filaments. The
relative axial motion of a filament pair $(l,k)$ at a given point of the
backbone is described by the crosslink shear displacement, which is the sum of a
geometric contribution, $\Delta_{lk}$, and the relative stretching of
neighboring filaments, $\Delta u_{lk}= u_l -u_k$. The geometric part results
from the arclength mismatch between the two filaments, induced by a bending and
twisting of the bundle central line. As in Eq.~(\ref{eq:Hstretch}) we first
write the shear energy as a sum over all crosslink positions $s_i$ and then
perform the continuum limit, to get
\begin{eqnarray}\label{eq:Hshear} 
  H_{\rm shear}^{lk} &=& \frac{k_\times}{2}\sum_i\left[\Delta_{lk}(s_i) +\Delta u_{lk}(s_i)\right]^2\nonumber\\
  &\to&\frac{k_\times}{2\delta}\int_0^Lds \left( \Delta_{lk} +\Delta u_{lk}
  \right)^2\,,
\end{eqnarray}
where $k_\times$ is the shear stiffness of the individual crosslink.

For any bundle deformation, the associated value of $\Delta_{lk}$ can be
compensated for by stretching the filaments, making the shear energy vanish when
$\Delta u_{lk} = -\Delta_{lk}$.  At the same time, however, this would increase
the stretching energy, which may be unfavorable if the stretching stiffness
$k_s$ is rather large. For deformations on the scale of the bundle length
($u'\sim u/L$) the ratio of both energies gives the important parameter
$\alpha=k_\times L^2/k_s\delta^2$, which quantifies the relative strength of
both deformation modes~\cite{batheBPJ2008}.

As a final ingredient to the model we need to calculate the dependence of
$\Delta_{lk}$ on the bundle curvatures, $\Omega_\alpha$. Without going into the
details of an explicit derivation, we here just give the resulting expression.
For more details we refer the reader to Appendix~\ref{sec:deriv-shear-deform}.
The special case relevant for the microtubule geometry is also dealt with in
some detail in~\cite{hilfinger2008PhysBiol}.  To linear order in $\Omega_\alpha$
we find
\begin{eqnarray}\label{eq:Delta_mismatch} 
  \Delta_{lk} &=& b_{lk}\cos\alpha_{lk}
  \left(y_{lk}\Omega_3 - \int_0^sdt\Omega_2(t)\right) \\\nonumber 
  &&- b_{lk}\sin\alpha_{lk}\left(z_{lk}\Omega_3 - \int_0^sdt\Omega_1(t)\right)\,,
\end{eqnarray} 
where we defined $b_{lk}=|\mbf{b}_{lk}|$ as the distance between the
filament pair and $\alpha_{lk}$ as the angle of $\mbf{b}_{lk}$ with
respect to the $z$-axis.  Furthermore, $(y_{lk},z_{lk})$ are defined
as the cross-sectional coordinates of the \emph{midpoint} between the
filament pair.

In contrast to $H_0$, which is an expansion in the generalized curvatures
$\Omega_\alpha$, the shear energy $H_{\rm shear}$ is a function of the
\emph{integrated} curvatures, since $\Delta\sim b\int_s\Omega\sim bL\Omega$. For
terms beyond the harmonic contribution to be negligible, one has to assume that
the shear displacement is sufficiently small, $\Delta \ll a$, where $a$ is some
microscopic length-scale related to the size of the crosslink. In terms of the
bundle curvatures this implies $L\Omega\ll a/R\sim 1$, which is much more
restrictive than the range $R\Omega\ll 1$, over which $H_0$ can be approximated
by a harmonic form. As a consequence the bundle is only allowed to make small
excursions from its initial state, an assumption which is usually well satisfied
in bundles of stiff polymers like actin or microtubules, but certainly breaks
down under extreme loading conditions (e.g. to describe post-buckling) or in
more flexible objects like DNA.

With this ``weakly-bending'' assumption we reformulate the generalized
curvatures in terms of the lab-frame Euler
angles~\cite{alimEPJE2007,goldsteinBOOK}
\begin{eqnarray}
 \Omega_1 & = & \deriv{\phi}{s}\sin\psi\sin\theta + \deriv{\theta}{s}\cos\psi\nonumber\\
 \Omega_2 & = & \deriv{\phi}{s}\cos\psi\sin\theta - \deriv{\theta}{s}\sin\psi\nonumber\\ 
 \Omega_3 & = & \deriv{\psi}{s} + \deriv{\phi}{s}\cos\theta\nonumber\,.
\end{eqnarray}
As reference state we take $\phi_0=\pi/2$, $\theta_0=\pi/2$ and
$\psi_0=s\omega_0$, which corresponds to a straight, but pre-twisted bundle that
points along the $x$-axis. For small excursions around this reference state we
can linearize the equations such that
\begin{eqnarray}\label{eq:Omega} \Omega_1 & = &
\deriv{\phi}{s}\sin\psi_0 + \deriv{\theta}{s}\cos\psi_0\\ \Omega_2 & = &
\deriv{\phi}{s}\cos\psi_0 - \deriv{\theta}{s}\sin\psi_0\nonumber\\ \Omega_3 & = &
\deriv{\psi}{s}\nonumber\,,
\end{eqnarray}
and the angles are now measured relative to the reference state. As expected the
pretwist $\psi_0$ leads to a coupling of the angles $\theta$ and $\phi$. In the
following we are primarily concerned with the case of vanishing pretwist. Then,
the coupling terms vanish and we can simply set
\[(\Omega_1,\Omega_2,\Omega_3)=(\theta',\phi',\psi')\,.\] We will come back to
the case of pretwist in Section~\ref{sec:pretwisted-bundles}.

\subsection{Examples for the arclength mismatch $\Delta$}\label{sec:examples}

For the purpose of illustration we provide some examples of how the shear
displacement $\Delta$ depends on the geometry of the bundle cross-section and
the deformation of its central line.

\begin{figure}[h]
 \begin{center}
 \includegraphics[width=0.8\columnwidth]{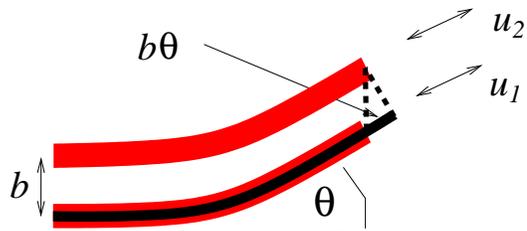}
\end{center} \caption{Illustration of crosslink shear deformation for the case
  of a two-filament bundle (filaments in red). Bundle deflection through the
  angle $\theta$ leads to the arclength mismatch, $\Delta=b\theta$. The
  filaments have to stretch the relative amount $u_1-u_{2} = b\theta$, in order
  to keep the crosslink (dashed line) undeformed with zero shear
  energy.}\label{fig:delta_two_filaments}
\end{figure}

If the bundle consists of only two filaments~\cite{everaers95} (geometry of a
ribbon) we have $y=z=\alpha=\pi/2$ as the central line and the mid-line between
the two filaments are identical. In this case $\Delta$ simplifies to
\begin{equation}\label{eq:Delta_two} \Delta = b\theta(s)\,,
\end{equation}
which is illustrated in Fig.~\ref{fig:delta_two_filaments}. Note, that twist
($\Omega_3$) does not contribute, as both filaments twist around the central
line symmetrically.

\begin{figure}[h]
 \begin{center}
 \includegraphics[width=0.8\columnwidth]{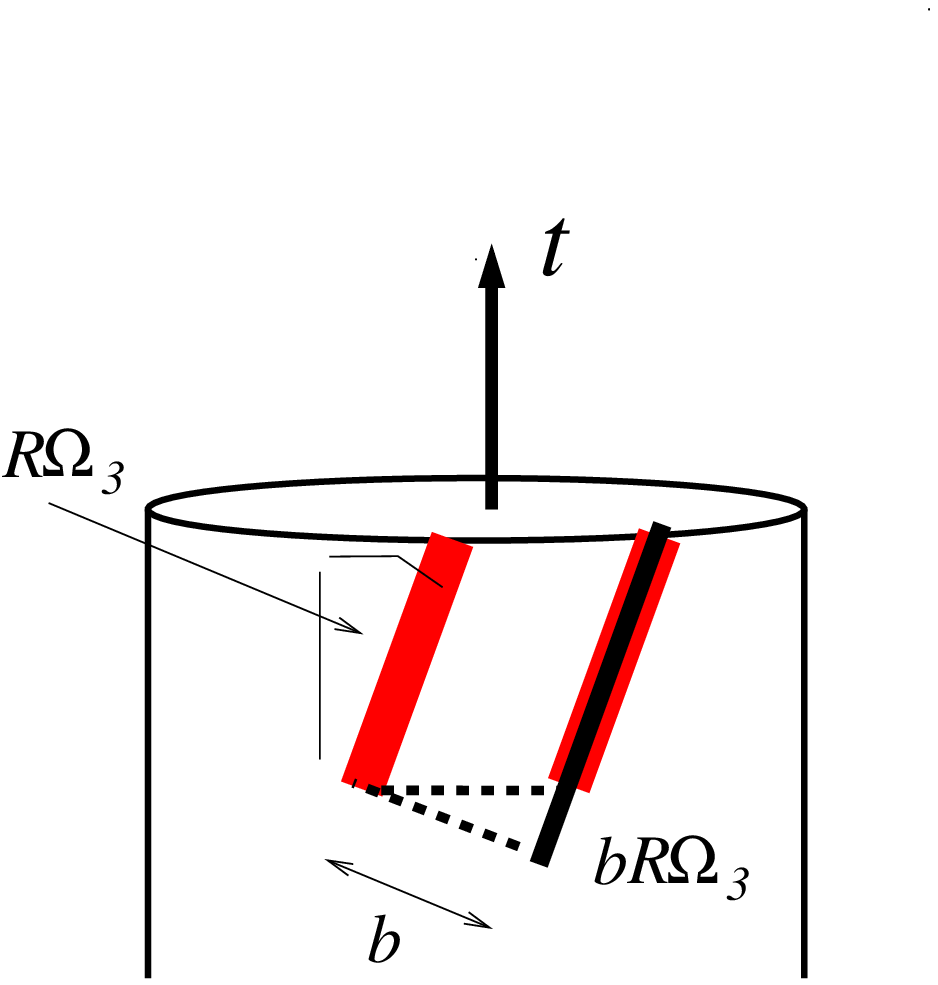}
\end{center} \caption{Illustration of crosslink shear deformation for a twisted
  microtubule. Indicated (in red) is a filament pair that winds around the
  microtubule cylinder (radius $R$) taking an angle $R\Omega_3$ with the
  cylinder axis $\mbf{t}$. The resulting arclength mismatch is given by
  $\Delta=bR\Omega_3$.}\label{fig:delta_mt}
\end{figure}

A particular case of this two-filament bundle has been considered in a set of
articles \cite{mergell2002PRE,golestanian2000PRE,Zhou2000PRE}, where the
crosslinks are assumed to be rigid with respect to shear, i.e.
$k_\times\to\infty$. To satisfy a vanishing arclength mismatch $\Delta$ one thus
requires $\Omega_1\equiv -\mbf{d}_2'\cdot \mbf{t} = 0$.  This means that the
unit vector $\mbf{d}_2$, which points from one filament to the other, must not
rotate in the direction of the tangent $\mbf t$. In other words, $\mbf{d}_2$
must equal the the bi-normal.  Thus the ribbon orientation is completely
specified by the space curve traced by the central line.

A second example is the axoneme in eukaryotic
flagellae~\cite{hilfinger2008PhysBiol}. There, filaments (microtubule doublets)
are arranged on the surface of a cylinder, just as the protofilaments in a
single microtubule. Switching to polar coordinates in the cross-sectional plane
(see Fig.~\ref{fig:quantities_explained}), $(y,z)\to(R,\varphi)$,
$\alpha=\varphi+\pi/2$, we find
\begin{equation}\label{eq:Delta_MT} 
  \Delta = -bR\psi' +  b\phi\sin\varphi+b\theta\cos\varphi\,.
\end{equation}
A similar expression, disregarding the possibility of twisting, has been given
by Mohrbach {\it et al.}~\cite{mohrbachPRL2007}. The structure of the second and
third term is the same as in Eq.~(\ref{eq:Delta_two}), additionally taking into
account the modified orientation in the cross-sectional plane, as described by
the angle $\varphi$. The origin of the first term is illustrated in
Fig.~\ref{fig:delta_mt} and elaborated on in
Appendix~\ref{sec:deriv-shear-deform}.

Consider now the case of planar bending. This will make the connection to
continuum elasticity particularly clear. Under planar deformation the bundle is
described by the one variable $\theta'=\Omega$ such that the shear deformation
is simply $\Delta = b\theta(s)\sim b\partial_x u_y$. Here, $u_y$ is the
displacement of the bundle transverse to the bundle axis (in the y-direction).
This latter form makes clear that the shear deformation represents one part of
the strain tensor component
$\epsilon_{xy}=\frac{1}{2}(\partial_xu_y+\partial_yu_x )$. The second part,
$\partial_yu_x\sim b\Delta u_{lk}$ is the continuum version of relative filament
stretching (see Eq.~(\ref{eq:Hshear})).

The elastic symmetries relevant to the WLB model depend on the arrangement of
the filaments in the cross-section. If there is rotational symmetry with respect
to the bundle axis, one speaks of transversly isotropic elastic
bodies~\cite{loveBOOK}. While, in general, this has five first order elastic
constants, our model has only two, augmented by the (second order)
bending/twisting elasticity of the individual filaments which is not accounted
for in continuum elasticity. The simplification arises from assuming transverse
inextensibility as well neglecting cross-sectional shape changes. The latter is,
for example, important in the failure of hollow tubes under bending.  One
relevant effect is the Brazier effect~\cite{calladineShellBook1983}, which
describes the increasing ovalization of the cross-section under the action of a
bending moment. In the present formulation of the model these nonlinearities are
not accounted for. For a discussion of potential modifications to include
cross-sectional deformations, we refer the reader to the outlook section at the
end of this article.

\subsection{Definition of effective bending and twist
  rigidities}\label{sec:effect-bend-twist}

It should be clear from the way the Hamiltonian was derived that the model is
applicable to bundles with arbitrary (ordered/disordered) arrangements of
filaments in the cross-section. In the remainder of this article we will focus
our attention to bundles with highly symmetric cross-sections, where the
filaments either form a rectangular array or a hollow tube.

In view of recent experiments probing the mechanical or statistical properties
of individual bundles in vitro~\cite{Claessens2006}, we head at a description of
the bundle in terms of effective bending and twist rigidities. These are defined
with respect to the standard worm-like chain model. To arrive at the proper
expressions we have to integrate out the internal stretching variable $u$, which
in general is not observable in experiment.

To show how this works, we symbolically write the partition function as $Z =
\int D\boldsymbol{\phi} Z(\boldsymbol\phi)$ where $\boldsymbol\phi$ signifies
the set of Euler angles $\phi(s),\theta(s),\psi(s)$. The constrained partition
function then reads $Z(\boldsymbol\phi) = \int D\{u\} \exp(-\beta
H(\{\boldsymbol\phi,u\}))\equiv \exp(-\beta W(\boldsymbol\phi))$.

The integration over the $u$-variables can easily be performed. As the
Hamiltonian is harmonic we are left with only Gaussian integrals, which are
evaluated in Fourier space. The resulting potential of mean force
$W(\boldsymbol\phi)$ can be written in the form of a wormlike chain Hamiltonian
\begin{equation}\label{def_eff_rigidities}
  W(\boldsymbol\phi_n) = \frac{L}{4}\sum_n q_n^2\left [\kappa_{B}(n)
    (\phi_n^2+\theta_n^2) + \kappa_{T}(n)\psi_n^2\right ] 
\end{equation}
with effective bending and twist rigidities $\kappa_{B}(n)$ and $\kappa_{T}(n)$,
respectively. We note that in the symmetric situations considered here there is
no coupling between the different deformation modes bending and twisting (see
Appendix~\ref{sec:exact-solution-2d}). In contrast to the usual WLC, the
effective bend and twist rigidities are in general dependent on the
mode-number $n$ and thus on the wavelength of the deformation. This effect and
the discussion of its consequences is the central topic of the remaining
sections.

\section{F-Actin bundle architecture}\label{sec:f-actin-bundle}

In the following sections we will focus our attention to bundles with
$N=(2M)^2$filaments that form a rectangular array (see
Fig.~\ref{fig:quantities_explained}). The angle $\alpha$ that specifies the
orientation of the filament-pair in the cross-section is then $\alpha=0,\pi/2$
as filaments are either arranged along the $y$- or the $z$-axis. As mentioned
above the different deformation modes decouple in harmonic order.  We can thus
investigate bending independently from twisting. Also the two space-directions
decouple and we can reduce the model to an effective two-dimensional
description~\cite{heussingerWLB2007}.

\subsection{Effective bending rigidity}\label{sec:effect-bend-stiffn}

The shear Hamiltonian reduces to
\begin{equation}\label{eq:HbendingOnly}
  H_{\rm shear} = \frac{Mk_\times}{\delta}\int_0^L\!ds
  \sum_{k=-M+1}^{M-1}(u_{k+1}-u_k+b\theta)^2 \,,
\end{equation}
where we used Eq.~(\ref{eq:Delta_mismatch}) with $\phi=\psi=0$. By following the
recipe outlined above we eliminate the axial strain variable $u_k$. By
approximating $u_k$ by a linearly increasing function of $k$ (see discussion
below) we arrive at the following result for the effective bending
rigidity~\footnote{A similar expression holds for bundles with hexagonal
  symmetry. For example, the factor $12$ should be substituted by $12N/(N+1/5)$.
  Also the length-scale $\lambda$ acquires a more complex (but essentially
  equivalent) dependence on bundle-size $N$.} as defined in
Eq.~(\ref{def_eff_rigidities})
\begin{equation}\label{eq:kappaLINEAR}
  \kappa_{B}(n) = N\kappa_b\left[ 1 + \left( \frac{12\hat\kappa_b}{N-1} +
      (q_n\lambda)^2   \right)^{-1} \right]\,.  
\end{equation}
Here, we have defined a characteristic wavelength
\begin{equation}\label{eq:lambda_square}
\lambda =\sqrt{\frac{2M}{(2M-1)}} \cdot \sqrt{\frac{\kappa_{b} \delta}{k_\times b^{2}}}\,,
\end{equation}
and a dimensionless bending rigidity $\hat\kappa_b=\kappa_b/(k_s\delta b^2)$. In
terms of the quantities $\lambda$ and $\hat\kappa_b$ the previously defined
$\alpha=k_\times L^2/k_s\delta^2$ can be rewritten as $\alpha\sim
\hat\kappa_b(L/\lambda)^2$. If the filaments behave as homogeneous elastic
beams, $\hat\kappa_b$ is just a number independent of bundle geometry or
crosslink spacing. For any numeric computation we will, for specifity, assume
that $\hat\kappa_b=1/12$, which corrresponds to beams with square
cross-sections~\cite{landauBOOKelasticity}.

\begin{figure}[h]
	\begin{center}
	\includegraphics[width=0.6\columnwidth,angle=-90]{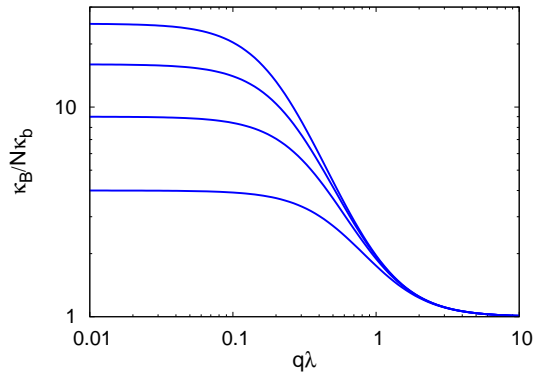}
	\caption{Effective bending rigidity, Eq.~(\ref{eq:kappaLINEAR}), as a
          function of mode-number, $q\lambda$, and for the set of bundle sizes
          $N=4,9,16,25$ (from bottom to top). In the fully-coupled and the
          decoupled regimes (corresponding to small and large $q$) the bending
          rigidity is constant.  At intermediate values of $q$ the bending
          rigidity scales as $\kappa_B(n)\sim k_\times q_n^{-2}$
          (shear-dominated regime).}
        \label{fig:kappaBend_q}
      \end{center}
\end{figure}

The characteristic feature of Eq.~(\ref{eq:kappaLINEAR}), is the wavelength
dependence (see Fig.~\ref{fig:kappaBend_q}). For wavelengths $q_n^{-1}$ in the
interval $1/\sqrt{N}\ll q_n\lambda \ll 1$ the bending stiffness decreases as
$\kappa_B(n)\sim k_\times q_n^{-2}$. In Ref.~\cite{batheBPJ2008} we have termed
this the intermediate or shear dominated regime as the bending rigidity is
proportional to the shear stiffness of the crosslinks. It is in this parameter
regime that the bundle behaves qualitatively different than either a homogeneous
beam (obtained in the "fully coupled" limit of $q_n\lambda\ll 1/\sqrt{N}$) or an
assembly of "decoupled" filaments ($q_n\lambda \gg 1$).

Another important feature of Eq.~(\ref{eq:kappaLINEAR}), which is independent
of the specific $q$-dependence, is the ratio of maximal to minimal bending
rigidity, $r=\kappa_{\max}/\kappa_{\min} = 1 + (N-1)/12\hat\kappa_b$. This only
depends on the number of filaments and the dimensionless bending rigidity
$\hat\kappa_b$.

\begin{figure}[h]
	\begin{center}
	\includegraphics[width=0.8\columnwidth]{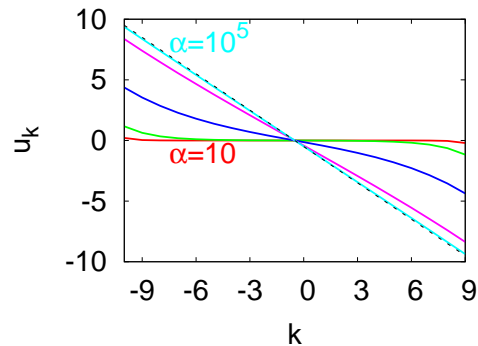}
	\caption{Dependence of axial strain $u_k$ on $k$, the distance from the
          bundle central line. Decreasing the dimensionless shear-stiffness
          $\alpha= k_\times L^2/k_s\delta^2$ the strain is reduced but not in a
          linear fashion. The outer layers of the bundle remain stretched
          stronger than the inner ones.}
	\label{fig:axialstrain} \end{center}
\end{figure}

While integrating out the stretching variables $u_k$ can be performed exactly,
Eq.~(\ref{eq:kappaLINEAR}) is based on the additional assumption that axial
strains are linearly increasing through the cross-section, $u_k =\Delta
u\cdot(k+1/2)$. The exact profile for $u_k$ is calculated in the appendix and
displayed in Fig.~(\ref{fig:axialstrain}); compared to the linear profile it
shows an enhancement of strain towards the bundle periphery.

\begin{figure}[h]
	\begin{center}
	\includegraphics[width=0.6\columnwidth,angle=-90]{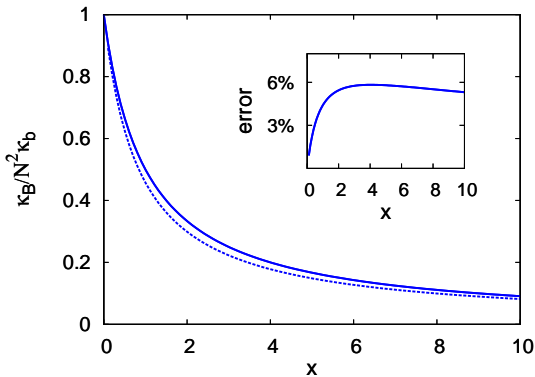}
	\caption{Continuum limit of effective bending rigidities as a function
          of $x=(q_nD)^2E/12G$.  Comparison of exact solution,
          Eq.~(\ref{eq:kappaContinuum}), with approximation,
          Eq.~(\ref{eq:kappaTIMO}). Inset: Relative error between both
          expressions.  The approximation over-estimates the bending stiffness
          by maximal $\sim 6\%$.}\label{fig:timo.wlb}
	\end{center}
\end{figure}

However, the ensuing value for the bending stiffness is largely insensitive to
the linear approximation~\footnote{See Fig.~\ref{fig:timo.wlb} for a discussion
  of the error in the continuum limit. In general, the error is smaller for
  smaller bundles and vanishes for a bundle consisting of only two filaments.}.
We speculate that the nonlinearities in the axial strain may eventually be
important for nonlinear material properties, as for example strain induced
rupture. The increased strain in the outermost filaments brings them closer to
their threshold for rupture and thus makes them more susceptible to this mode of
failure.

In order to make contact with continuum models for beam bending we perform a
continuum limit, by letting $N\to\infty$ but keeping the bundle aspect ratio
$D/L\sim bM/L$ constant. Then, bundle length $L$ has to grow with $M$ as
$L(M)\sim M$. In particular, this implies that fewer and fewer modes $n$ belong
to the decoupled regime (where $q_n^{-1} \ll \lambda $). Eventually, this
regime, where filaments bend independently ($\kappa_B(n)\approx N\kappa_b$) is
no longer accessible. In effect this means that the bending stiffness $\kappa_b$
of the individual filaments can be neglected, just as in "normal" continuum
elasticity, where higher order gradients (${\cal O}(\theta')$) are not accounted for
from the start.

In this continuum limit the result from the linearization assumption,
Eq.~(\ref{eq:kappaLINEAR}), reduces to the Timoshenko model for beam
bending~\cite{timoshenko}, which was recently used to interpret bending
stiffness measurements on microtubules~\cite{kis02,pampaloni06} and carbon
nanotube bundles~\cite{kis04},
\begin{equation}\label{eq:kappaTIMO}
  \kappa_B^{\rm TIMO}(n) = \frac{N^2\kappa_b}{1+(q_nD)^2E/12G}\,.
\end{equation}
To allow for comparison with continuum elasticity we have used the expressions
$k_s\delta=Eb^2$ and $\kappa_b=Eb^4/12$ applicable for homogeneous beams of
square cross-section and defined the shear-modulus $G=k_\times/\delta$.

On the other hand, one can equally derive a continuum limit from the exact
expression for $\kappa_{B}(n)$ (as presented in Appendix~\ref{sec:bending}). This
gives
\begin{equation}\label{eq:kappaContinuum}
  \kappa_{B}^{\rm CONT}(n) = \frac{N^2 \kappa_b}{(q_nD)^2E/12G}\left(
    1-\frac{\tanh(q_nD\sqrt{E/4G})}{q_nD\sqrt{E/4G}}
  \right)\,.
\end{equation}

Both expressions, Eq.~(\ref{eq:kappaTIMO}) and Eq.~(\ref{eq:kappaContinuum}) are
compared in Fig.~(\ref{fig:timo.wlb}). One infers that the approximation
(Timoshenko theory) overestimates the exact solution by no more than $6\%$. The
difference can partly be compensated for by introducing a shear-correction
factor ($\approx1.2$) in the denominator of Eq.~(\ref{eq:kappaTIMO}).

\subsection{Tangent-tangent correlation function}\label{sec:tang-tang-corr}

In this section the implications of a mode-number dependent bending rigidity is
further elaborated by discussing the concept of the persistence length. The
persistence length of a single WLC may be defined in terms of the competition of
bending and thermal energies, $l_p=\kappa/k_BT$.  With this definition, bending
rigidity and persistence length are basically identical. In the framework of the
WLB this would lead to a mode-number dependent persistence length
$l_{p}(n)=\kappa_{B}(n)/k_BT$.

For the WLC the persistence length is also the length-scale over which the
tangent-tangent correlation function decays
\begin{equation}\label{eq:tagengt-tangent}
  \langle \mbf t(s)\mbf t(0) \rangle = \exp\left[-s/l_p\right]\,.
\end{equation}

This simple exponential form is no longer valid for the WLB as can be
illustrated by considering planar bending with $\phi=0$. Then the
tangent-tangent correlation function is easily inferred from the angular
fluctuations~\cite{everaers95} as
\begin{equation}\label{eq:}
  \langle \mbf t(s)\mbf t(0) \rangle = \exp\left[
    -\frac{1}{2}\left\langle\left(\theta(s)-\theta(0) \right)^2\right\rangle
  \right] \,,
\end{equation}
with
\begin{equation}\label{eq:lp_effective}
  \left\langle\left(\theta(s)-\theta(0) \right)^2\right\rangle  =
  \frac{1}{Nl_p}\left[ \frac{A-1}{A}s + \frac{\D\lambda}{\D
      A\sqrt{A}}\left(1-e^{-s\sqrt{A}/\lambda}\right)    \right]   \,,
\end{equation}
and $A = 1+12\hat\kappa_b/(N-1)$. Forcing such a complex expression into the
form given by Eq.~(\ref{eq:tagengt-tangent}) implies an arclength-dependent
persistence length, $l_p(s)$. At short distances, one recovers the decoupled
regime and $l_p(s)=Nl_p$, while at long distances, $l_p(s)=N^2l_p$ as found in
the fully coupled regime.

Note, that there is no immediate relation between this $l_p(s)$ and the $l_p(n)$
defined above. The Fourier-transform of $l_p(n)$ is, instead, given by the
following expression
\begin{equation}\label{eq:}
  l_p^\star(s) = \frac{N\kappa_b}{k_BT}\left( L\delta(s) +
    \frac{L}{\lambda}\frac{e^{-\sqrt{A-1}s/\lambda}}{\sqrt{A-1}} \right)\,.
\end{equation}
This quantity appears in the elastic energy expressed in real-space as 
\begin{equation}\label{eq:Hnonlocal}
  H_{\rm WLB} = k_BT\int\theta'(s_1)l_p^\star(s_1-s_2)\theta'(s_2)ds_1ds_2\,, 
\end{equation}
which is a non-local function of arclength. The length-dependence obtained here
is markedly different from that found in the correlation function,
Eq.~(\ref{eq:lp_effective}). While $l_p(s)$ is constant at large distances,
$l_p^\star(s)$ decays exponentially and vanishes over the length-scale
$\lambda/\sqrt{A-1}\sim\lambda\sqrt{N/\hat\kappa_b}$, which corresponds to the
onset of the fully-coupled regime.

A similar nonlocal energy function is obtained when considering the fluctuation
properties of elastic membranes. In-plane shear- and compression-modes lead to a
renormalized bending rigidity for the out-of-plane
fluctuations~\cite{nelson87,nelsonBOOK}. In contrast to bundles, however, there
the kernel is long-ranged $\l_p(s)\sim 1/s$, which asymptotically leads to a
flat membrane phase.

Concluding this section, we find that it is impossible to speak of a single
persistence length without specifying the precise experimental conditions as
well as the observable under consideration (here tangent-tangent correlation
function). The WLB model presents a framework within which a length-dependent
bending stiffness can be understood. However, we would like to emphasize that
the fundamental quantity is the $q_n$-dependent $\kappa_B(n)$ as presented in
Eq.~(\ref{eq:kappaLINEAR}) or below in Eq.~(\ref{eq:kappaMT}). It depends on
wavenumber in a universal way independent of the type of measurement. The
dependence on bundle length, in contrast, arises through a specific
transformation to real space and may produce different results depending on the
observable under consideration.

\subsection{Frequency-dependent correlation and response
  functions}\label{sec:freq-depend-corr}

In our previous publications~\cite{batheBPJ2008,heussingerWLB2007} we have
calculated several thermodynamic observables and showed that the mode-dependent
bending stiffness of a WLB may lead to drastic modifications of their scaling
behavior. Here, we widen the scope of this analyis by discussing dynamic
observables. In analogy to the usual overdamped dynamics of a WLC we can discuss
the dynamics of a WLB by substituting the mode-dependent bending stiffness in
the standard Langevin equation of motion for the transverse bending modes
$r_\perp(q_n,t)$
\begin{equation}\label{eq:}
 \zeta \frac{\partial r_\perp}{\partial t} = -\kappa_B(n)q_n^4r_\perp+\xi(q_n,t)\,.
\end{equation}

With this one obtains for the reduced correlation function
\begin{eqnarray}\label{eq:corrFct}
  C(t) &:=& L^{-1}\int_0^L \left\langle \left(r_\perp(s,t)-r_\perp(s,0)\right)^2\right\rangle ds\,.\nonumber\\
 &=& \frac{k_BT}{L}\sum_n\frac{1-e^{-t/\tau_n}}{\kappa_B(n)q_n^4}\,,
\end{eqnarray}
with the relaxation times $\tau_{n}=\zeta/\kappa_B(n)q_n^4$.  For WLCs (constant
$\kappa_B\equiv\kappa$) one finds a scaling regime at times $t<\tau_1\sim \zeta
L^4/\kappa$, where $C(t)\sim t^{3/4}$ \cite{freynelson1991,kroyPRE1997,isa96}.

For WLBs, on the other hand, one has to use the $q$-dependent effective bending
stiffness, Eq.~(\ref{eq:kappaLINEAR}), instead. As the term $(q\lambda)^{-2}$
multiplies the $q^4$ contribution in Eq.~(\ref{eq:corrFct}) one finds the
correlations to grow in time as $C_{\rm WLB}(t)\sim t^{1/2}$. Of course, this
result is only valid as long as the $q\lambda$-term dominates the effective
bending stiffness, i.e. as long as the bundle is in the intermediate regime. In
fact, there will be a complex cross-over scenario
\begin{equation}\label{eq:dynamic_correlation}
  C(t) \sim \begin{cases}
    (Nl_p)^{-1}\left(\dfrac{N\kappa_bt}{\zeta}\right)^{3/4}\,, & \text{$t \ll
      t_1$} \\ 
    (Nl_p)^{-1} \lambda^2\left(\dfrac{N\kappa_bt}{\zeta}\right)^{1/2}\,,
    & t_1 \ll t \ll t_2 \\ 
    (Nl_p)^{-1}
    \left(\dfrac{\hat\kappa_b}{N}\right)^{1/4}\left(\dfrac{N\kappa_bt}{\zeta}\right)^{3/4}\,,
    & \text{$ t_2 \ll t \ll
      t_3$} \\

   (Nl_p)^{-1}L^3\dfrac{\hat\kappa_b}{N}\,, & \text{$t \gg t_3$}
   \end{cases}
\end{equation}
with the cross-over times $t_1=\zeta\lambda^4/N\kappa_b$ and
$t_2=t_1\dfrac{\hat\kappa_b}{N}$ governing the cross-over from the decoupled to
the intermediate and the fully-coupled regimes. At times larger than
$t_3=t_1\dfrac{\hat\kappa_b}{N}\dfrac{L^4}{\lambda^4}$ the correlation function
saturates.

{F}rom the correlation function one can furthermore calculate the response
function $\chi_\perp$ measuring the linear response to transverse forces.  Using
the fluctuation-dissipation theorem and the Kramers-Kronig relations one finds
\begin{equation}\label{eq:responsePerp}
  \chi_\perp(\omega) = \sum_n\frac{1}{L}\frac{1}{\kappa_B(n)q_n^4-i\zeta\omega}\,.
\end{equation}
This contrasts with the response function $\chi_\parallel$ for
stretching forces
\begin{equation}\label{eq:}
\chi_\parallel(\omega) = \sum_n\frac{1}{l_p(n)}\frac{1}{\kappa_B(n)q_n^4-i\zeta\omega/2}\,,
\end{equation}
which is sometimes taken as a starting point to determine the high-frequency
response of entangled solutions of stiff
polymers~\cite{gittesPRE1998,morsePRE1998,morseMacroTwo1998,granek97,koenderinkPRL2006},
The transverse response, on the other hand, has been argued to relate to a
``microrheological modulus''~\cite{freyBook2000,kroyNJP2007} that is measured
locally by imbedding probe beads into the network.  Unfortunately, for a WLC
both functions are hardly indistinguishable, and only differ by the constant
factor $\chi_\parallel/\chi_\perp = L/l_p$. The high-frequency behavior in both
cases is $\chi\sim \omega^{-3/4}$.

\begin{figure}[h]
	\begin{center}
	\includegraphics[width=0.8\columnwidth]{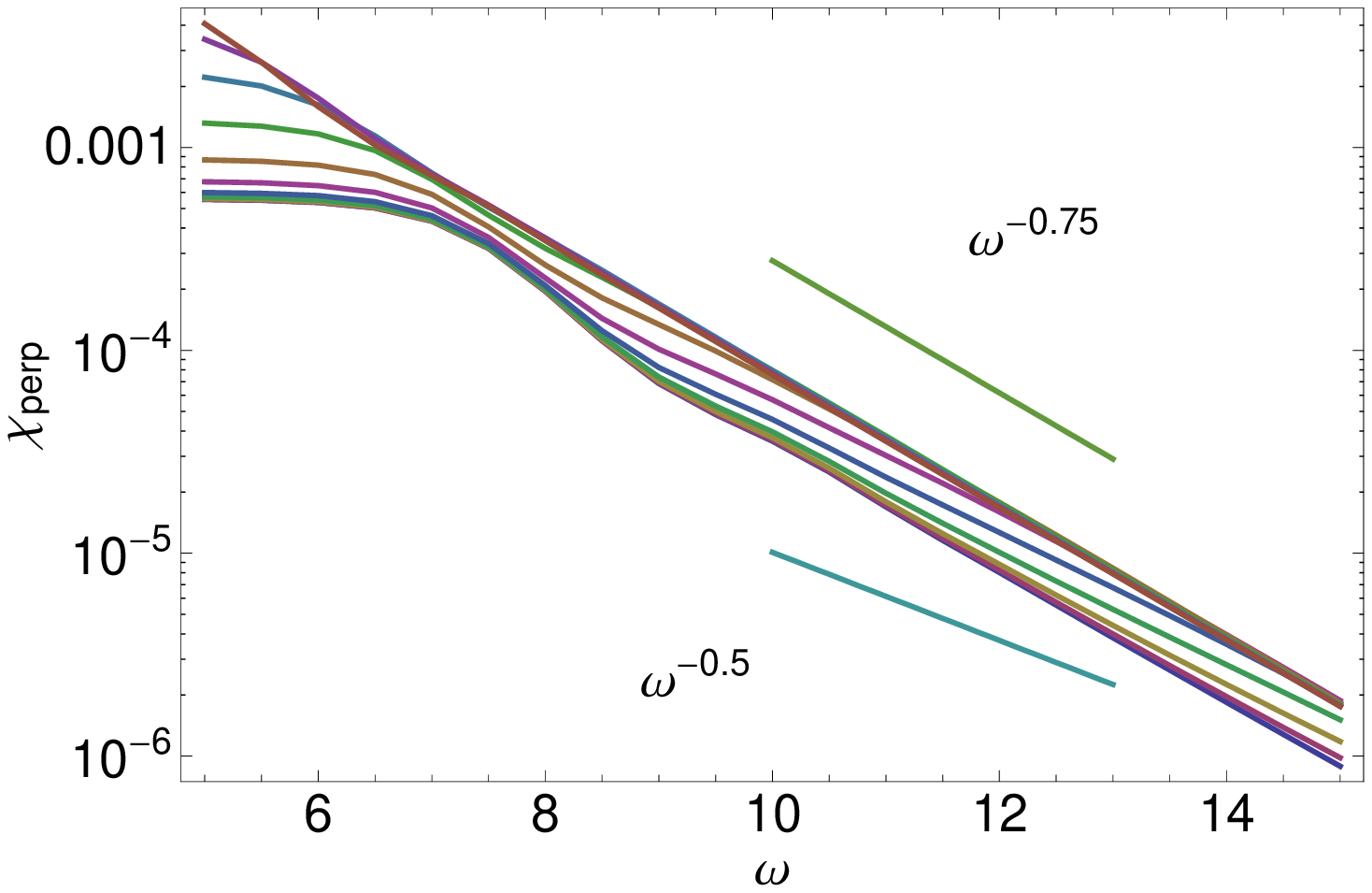}
        \hspace{1cm}
        \includegraphics[width=0.81\columnwidth]{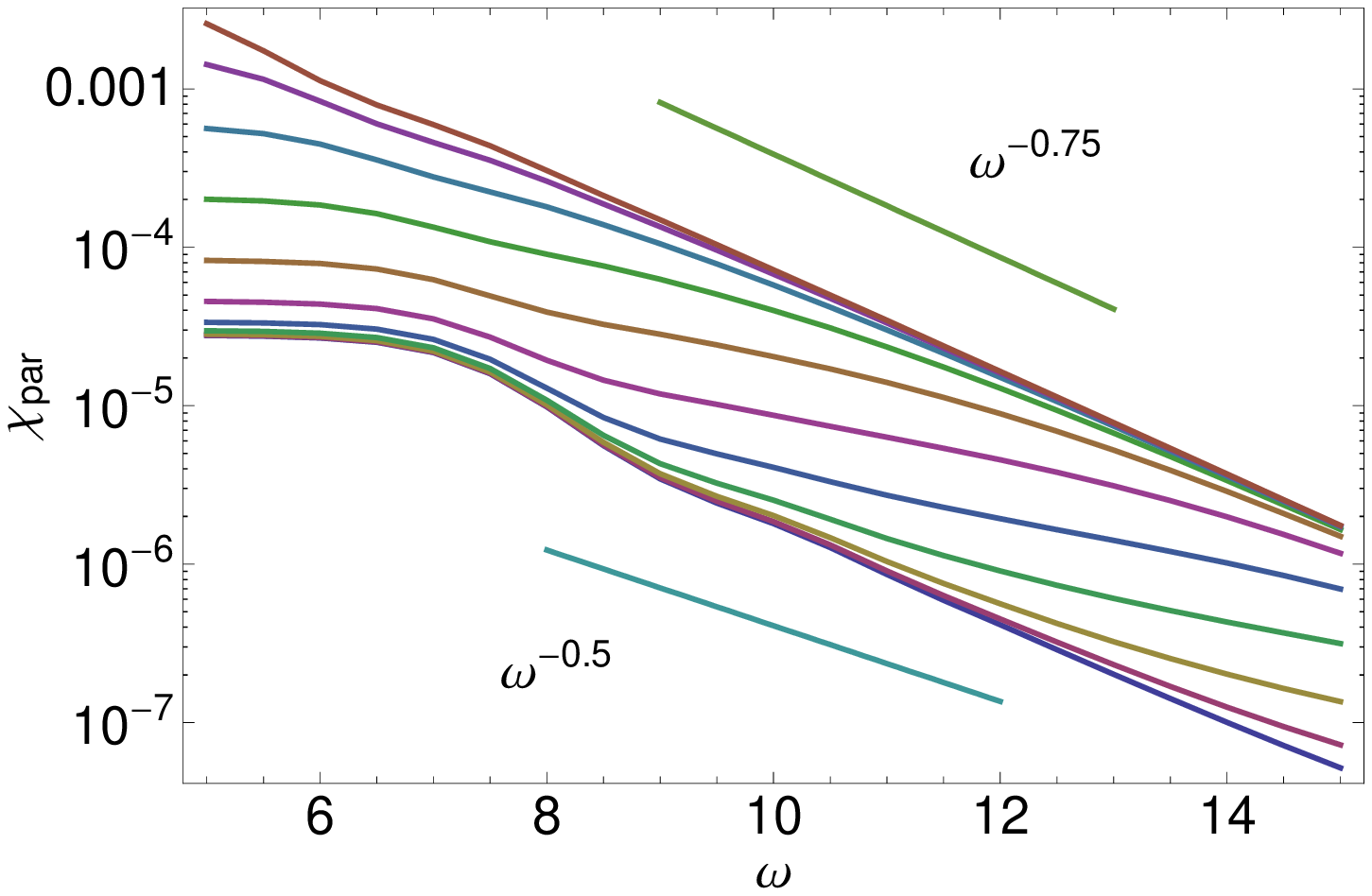}
	\caption{Comparison of transverse response function $\chi_\perp(\omega)$
          (top) with longitudinal response function $\chi_\parallel(\omega)$
          (bottom) for a bundle of $N=20$ filaments (in units of $N\kappa_b/L^3$
          and assuming $Lk_BT/N\kappa_b=1$).  Frequencies are plotted in units
          of $\zeta L^4/N\kappa_b$. The different curves correspond to different
          values of $\lambda/L$. The two asymptotic scaling regimes with
          $\omega^{-3/4}$ correspond to the decoupled (top) and fully-coupled
          regime (bottom), respectively. The intermediate $\omega^{-1/2}$ is
          sharper in $\chi_\perp$ than in $\chi_\parallel$ even though the
          latter function shows overall a stronger variation with frequency.}
        \label{fig:chiperppar} \end{center}
\end{figure}

In the case of a WLB, things are somewhat different, as the additional factor
$L/l_p(q)$ in the longitudinal response function not only changes the prefactor
but also modifies the functional form with respect to frequency (see Fig.~\ref{fig:chiperppar}).

For similar reasons as in the discussion of the correlation function one expects
an intermediate regime with $\omega^{-1/2}$, at least in the transverse
response. This should lead to measurable signatures in microrheological
experiments on bundled F-actin systems. Due to the additional $q$-dependence in
the denominator, the longitudinal response, $\chi_\parallel$ shows a smooth
cross-over between the two asymptotic regimes of fully-coupled and decoupled
bending, as explained in the figure.

\subsection{Effective twist rigidity}\label{sec:effect-twist-rigid}

We now turn to the discussion of the twist mode. In this case the Euler angles
$\phi$ and $\theta$ in Eq.~(\ref{eq:Delta_mismatch}) are zero such that the
shear Hamiltonian reduces to
\begin{equation}\label{eq:HtwistOnly}
H_{\rm shear} = \frac{k_\times}{2\delta}\int_0^L\!ds
    \sum_{lk}\left(\Delta u_{lk}+bd_{lk}\psi'\right	)^2 \,,
\end{equation}
where we defined the geometric factor $d_{lk} = y_{lk}\cos\alpha_{lk} -
z_{lk}\sin\alpha_{lk}$.

As may already be apparent from comparing Eq.~(\ref{eq:HbendingOnly}) with
Eq.~(\ref{eq:HtwistOnly}) the stretching deformation $u$ couples differently to
twist ($k_\times u\psi'$) as to bending ($k_\times u\theta$). The difference
being the additional derivative occuring in Eq.~(\ref{eq:HtwistOnly}).
Effectively this means that the resulting twist rigidity will not depend on
mode-number $q_n$ but receive a constant correction to the single filament value
$\kappa_{t}$.

Let us first assume that the filaments are inextensible. Then the $u$-terms in
the Hamiltonian identically vanish and the effective twist rigidity can simply
be read off from the terms multiplying $\psi'^2$,
\begin{eqnarray}\label{eq:kappaTwist}
 \kappa_{T} &:=& N\kappa_{t} \left[ 1 + \frac{N-1}{6} \left( \frac{\lambda_t}{b} \right)^{-2} \right]\,.
\end{eqnarray}
Here we have defined $\lambda_t := \sqrt{2M/(2M-1)} \sqrt{\kappa_{t} \delta /
  k_{\times} b^{2}}$, which is similar to $\lambda$ defined above, with
$\kappa_b$ substituted by $\kappa_t$. Unlike $\lambda$, however, it multiplies
the constant length $b$, the lateral distance between neighboring filaments. As
anticipated, there is no mode-number dependence.

By introducing bundle diameter $D$ and effective shear modulus $G$ (as in the
discussion leading to Eq.~(\ref{eq:kappaTIMO})) the second term takes a form
well known from continuum theory, $\sim GD^4$~\cite{landauBOOKelasticity}. Thus,
we find an effective twist rigidity, Eq.~(\ref{eq:kappaTwist}), that just
describes the simple additive superposition of two contributions, shear-induced
rigidity (second term) and twist rigidity of the individual filaments (first
term). In the continuum limit $N\to\infty$ the latter contribution can naturally
be neglected as it only grows with $N$ as compared to the $N^2$ increase of the
shear-induced rigidity.

\begin{figure}[h]
	\begin{center}
	\includegraphics[width=0.8\columnwidth]{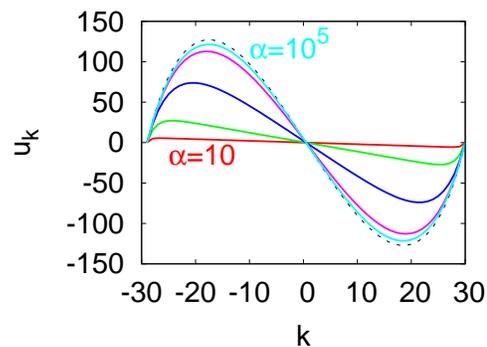}
	\caption{Cross-sectional warping $u_{k}$ ($k=-M+1\ldots M$,
          corresponding to one row of the rectangular array) for different
          non-dimensional crosslink shear stiffness $\alpha=k_\times
          L^2/k_s\delta^2$. The classic solution for a beam (dashed line) is
          approached as $k_\times\to\infty$. In the opposite limite,
          $k_\times\to0$, there is no warping and $u\equiv0$, as filaments
          remain uncoupled in this limit.}
        \label{fig:cont_limit_twist} \end{center}
\end{figure}

Now we allow for finite axial displacements $u_k$. This is commonly referred to
as cross-sectional warping; under twist deformations the bundle cross-sections
do not stay plane but deform and acquire a curvature. Just as in the case of the
bending rigidity, the exact solution for the twist rigidity only differs little
from the foregoing simplified analysis. Some details about the derivation are
presented in Appendix~\ref{sec:twisting}.  The resulting axial displacements
$u_k$ can be found in Fig.~(\ref{fig:cont_limit_twist}) The classic solution of
Saint-Venant~\cite{loveBOOK} is approached closer and closer for increasing the
shear-stiffness $k_\times$.

\section{Microtubule architecture}\label{sec:micr-nanot}

In this section we want to turn our attention to the case of microtubules, which
we model as bundles with filaments arranged on the surface of a cylinder.  For
the time being we assume that in the groundstate the microtubule is untwisted
such that the $N$ protofilaments are oriented along the cylinder axis. This
assumption is in fact only valid for microtubules with $N=13$
protofilaments~\cite{chretien1992JCellBio,chretien1991BiolCell}. This class,
nevertheless, seems to be the most frequent in in-vitro polymerization
experiments.

There is an ongoing debate in the literature about the dependence of microtubule
bending rigidity on
length~\cite{kurachiCellMotCytosk1995,kikumotoBPJ2006,brangwynne2007BPJ,pampaloni06,taute2008PRL,heuvel2008PNAS,heuvel2007NanoLett,kis02}.
Early buckling experiments indicated a
length-dependence~\cite{kurachiCellMotCytosk1995}, while an improved version of
the same experiments later gave a negative result~\cite{kikumotoBPJ2006}.
Brangwynne {\it et al.}  \cite{brangwynne2007BPJ} performed a mode analysis of
microtubule contours. Their data is compatible with a length-dependent bending
rigidity but the authors vote for a cautious interpretation of the results in
view of the close proximity to the noise level. Recently, experiments by
Pampaloni {\it et al.} \cite{pampaloni06} measured the transverse fluctuations
of grafted microtubules to establish an increasing persistence-length for
microtubule lengths up to $L\sim 23\mu m$.  Using a high-precision tracer
technique, Taute {\it et al.}  \cite{taute2008PRL} analyzed shorter microtubules
and found the persistence length to level off at $l_p^{\rm min}\approx 580\mu m$
for lengths shorter than $L\approx 5\mu m$. Similar values have been obtained in
Ref.~\cite{heuvel2008PNAS} ($l_p^{\rm min}=90\mu$m) and in
Ref.~\cite{heuvel2007NanoLett} ($l_p^{\rm min}=240\mu$m), and explained with the
help of the WLB model. On short length-scales (in the decoupled regime) the
effective bending rigidity is constant because it reflects the stiffness of the
individual proto-filaments. In contrast, Kis {\it et al.}~\cite{kis02} have
found a decreasing bending stiffness even for microtubule sections as short as
several hundred nanometers.

While the discussion about these partially conflicting measurements is ongoing,
we would like to point out that different techniques do not necessarily have to
come to the same conclusion, once the idea of the bending rigidity as a
fundamental material parameter is given up. In the context of the WLB model the
fundamental quantity is a mode-dependent bending rigidity. As discussed in
Section \ref{sec:tang-tang-corr}, any dependence on bundle length is a
``secondary'' effect that will depend on the observable under consideration.

By using the same procedure as in the case of the rectangular bundle we find for
the microtubular bending rigidity
\begin{equation}\label{eq:kappaMT}
  \kappa_{B}(n) =N\kappa_b\left [1+
    \left(\frac{8\hat\kappa_b}{\sin^{-2}(\pi/N)}+(q_n\lambda)^2\right)^{-1}
  \right]\,,
\end{equation}
where the relevant length-scale $\lambda$ is now defined as
$\lambda=\sqrt{\kappa_{b} \delta / k_\times b^{2}}$.  Note, that this expression
results in the bending stiffness in the fully-coupled regime (where the
microtubule behaves as a simple beam) to scale as $\kappa\sim N^3$, in contrast
to bundles with a homogeneous cross-section (as the one discussed before), where
$\kappa\sim N^2$.

With Eq.~(\ref{eq:kappaMT}) direct comparison with experimental data can be
made. In particular, the ratio of maximal to minimal bending rigidity can most
easily be determined, $r=1+\sin^{-2}(\pi/N)/8\hat\kappa_b=1+2/\sin^2(\pi/N)$.
For the latter equality we assumed the protofilament to have a circular
cross-section ($\hat\kappa_b=1/16$).  For microtubules with $N=13$
protofilaments this results in a universal ratio $r\approx 35$.

In this way, using the the range of values $l_p^{\min}\approx0.1\ldots0.6$mm
(which we assume to represent the protofilament stiffness) one can estimate the
maximal persistence length of long microtubules to be in the range
$l_p^{\max}\approx 3.5\ldots21$mm. Compared with
experiments~\cite{jansonBPJ2004,gittesJCB1994,pampaloni06} these values seem to
be too large. Given the large error-bars in any of the mentioned experiments
this calculation may, nevertheless, be acceptable. Furthermore, as we will
discuss below, microtubule helicity can provide a mechanism to reduce the
apparent persistence length by reducing the effect of shear-induced coupling.
For given microtubule length the persistence length is then predicted to
decrease with increasing helicity -- thus improving the comparison with
experiment.

Finally, let's turn to the case of pure twist deformations. Due to the symmetry
of the circular cross-section no warping is possible.  We find for the
microtubule twist rigidity, similar to Eq.~(\ref{eq:kappaTwist}),
\begin{eqnarray}\label{eq:kappaTwistMT}
  \kappa_{T} &:=& \kappa_{t} \left[ 1 + \frac{\tan^{-2}(\pi/N)}{4} \left(
      \frac{\lambda_t}{b}\right)^{-2} \right]\,, 
\end{eqnarray}
where $\lambda_t := \sqrt{\kappa_{t} \delta / k_\times b^{2}}$.

\section{Pretwisted bundles}\label{sec:pretwisted-bundles}

In the previous sections we have restricted our attention to non-helical bundles
and assumed that in the ground-state the filaments point along the bundle axis.
As a final application of our model, we will here discuss the question of
helicity, or pretwist, and its influence on bundle mechanics. This aspect is
important not only for some types of microtubules but also for F-actin bundles
and is reflected in the role that helicity plays in providing an explanation for
the existence of a (thermodynamically) preferred bundle
size~\cite{grasonPRL2007,grasonPRE2009,claessensPNAS2008}.

The discussion of pretwisted bundles proceeds in two steps. We first assume the
bundle to form with all filaments straight. Starting from this reference state
crosslink binding may add a driving force for twisting the bundle if the
straight state does not allow for optimal accessibility of the crosslink
binding sites. Another effect may be the helicity of the filaments themselves
that favor a twisted bundle over an untwisted one.

Without elaborating on the precise mechanism that leads to pretwisted bundles,
we incorporate bundle pretwist by substituting the generalized twist-curvature
$\Omega_3$ by $\Omega_3-\omega_0$. In doing so the new energetic ground-state is
at $\Omega_1=\Omega_2=0$ and $\Omega_3=\omega_0$ . To obtain the effective
bending rigidity we linearize the curvatures around this ground-state, as
performed in Eq.~(\ref{eq:Omega}).

\begin{figure}
	\centering
	\includegraphics[width=0.8\columnwidth]{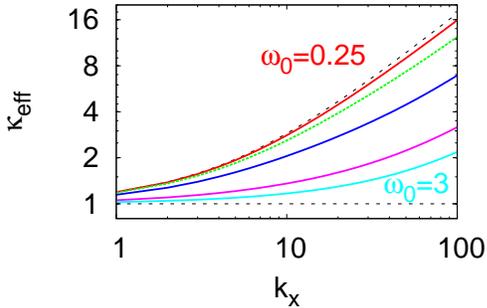}
	\caption{Effective bending stiffness (in units of $N\kappa_b$) as a
          function of shear stiffness $k_\times$ (in units of
          $\kappa_b\delta/b^2L^2$) of a bundle of four filaments under a tip
          force $F$. The bending stiffness is calculated from the determined
          end-deflection $y(L)$ by $\kappa_{\rm eff} = FL^3/12y(L)$. The full
          curves are for finite pretwist $\omega_0L/\pi=(0.25,0.5,1,2,3)$, while
          the dashed curves represent the limiting cases of zero and infinite
          pretwist, repsectively.  }\label{fig:pretwist}
\end{figure}

Inserting this result into the shear deformation, Eq.~(\ref{eq:Delta_mismatch}),
one finds terms like $b\int \cos(\psi_0)\phi'ds$ which depend nonlinearly on
arclength $s$. Thus a transformation to Fourier-space is no longer helpful as
different modes would remain coupled. We therefore resort to an alternative
approach, and determine the effective bending rigidity by numerical integration
of the mechanical equilibrium equations in real space, $\partial
H/\partial\boldsymbol{\phi} = 0$.  Specifically, we determine the end-deflection
$y(L)$ of a bundle of 4 filaments under a point force $F$ at the distal end
($s=L$), given clamped boundary conditions at the proximal end ($s=0$). The
effective bending rigidity is then obtained from the expression, $\kappa_{\rm
  eff} = FL^3/12y(L)$, and displayed in Fig.~(\ref{fig:pretwist}) as a function
of the crosslink shear stiffness, $k_\times$, and a series of values for the
pretwist, $\omega_0$. For simplicity we have assumed the filaments to be
inextensible, $k_s\to\infty$, and thus restricted ourselves to the decoupled and
the intermediate regimes.

For increasing pretwist the apparent stiffness decreases and asymptotically
approaches the value without shear-stiffness. Thus, in pretwisted, helical
bundles the crosslinks only have a limited ability to mechanically couple the
different filaments together. The higher the twist the more the filaments act as
if they were independent~\cite{everaers95}. The reason for this behavior is that
in pretwisted bundles the filaments exchange their place and those that start on
the top of the bundle soon are on the bottom. Thus, crosslink sites that would
stay behind and lead to large shear displacements in untwisted bundles, can now
catch up to make the effective shear deformation smaller.

\section{Conclusions and Outlook}\label{sec:conclusions-outlook}

We have presented a detailed study of the elastic and dynamic properties of
bundles of semiflexible filaments (wormlike bundle model, WLB). It is found that
a competition between the elastic properties of the filaments and those of the
crosslinks leads to renormalized effective bend and twist rigidities that can
become mode-number dependent. The strength and character of this dependence
varies with bundle architecture, such as the arrangement of filaments in the
cross section and pretwist.

Two paradigmatic cases of bundle architecture have been discussed (see
Fig.~\ref{fig:quantities_explained}): the first assumes filaments to be arranged
homogeneously throughout the cross-section, for example on a square or
triangular lattice. This geometry is particularly relevant for F-actin bundles.
The second architecture has the filaments arranged on the surface of a cylinder
as is the case for microtubules. For all bundle architectures, the bending
rigidity depends on mode-number $q_n$ as $\kappa_B(n)\sim k_\times q_n^{-2}$.
This is the shear dominated regime, as the bending rigidity is proportional to
the shear stiffness, $k_\times$, of the crosslinks~\cite{batheBPJ2008}. It is in
this parameter regime that the bundle behaves qualitatively different than
either a homogeneous beam (obtained in the "fully coupled" limit) or an assembly
of "decoupled" filaments. Each architecture has its own universal ratio of
maximal to minimal bending rigidity, independent of the specific type of
crosslink induced filament coupling. For microtubules (without pretwist) we find
the ratio $r=1+2/\sin^2(\pi/N)$, which is in reasonable agreement with the
available experimental data.

An important factor in determining the strength of crosslink-induced filament
coupling is the pretwist (helicity) of the bundle. Numerical computation shows
that the effective bending rigidity decreases with increasing the pretwist. This
has interesting consequences for microtubules, where the amount of pretwist
depends on the number of protofilaments, $N$. Different microtubule types are
therefore predicted to have different variations in their effective bending
rigidity. These predictions could be tested in experiments that are able to
select the microtubule type and measure their bending rigidity independently.

We have discussed several further observables, static and dynamic, that could be
relevant to experiments. We have shown that the concept of the persistence
length becomes ambiguous and depends on the observable used. The usual
definition via the tangent-tangent correlation function is shown to lead to a
persistence-length that depends on the scale of observation. Further observables
that are effected by the mode-dependent bending rigidity are the force-extension
relation or the buckling force~\cite{heussingerWLB2007}.  Interestingly, in the
intermediate regime the latter is constant and independent of bundle length.

The dynamic properties of bundles are characterized by a complex cross-over
scenario which is in one-to-one correspondence with the three regimes of
decoupled, intermediate and fully-coupled bending. While decoupled and
fully-coupled bending display the usual $t^{3/4}$ in the correlation function,
it is shown that the shear-coupling leads to an intermediate asymptotic regime,
where the correlations only grow as $t^{1/2}$. The response functions for
longitudinal and transverse forces also reflect these different regimes. In
contrast to the WLC, they are not just proportional to each other but show
distinct frequency dependences. These findings may be relevant for
microrheological experiments, with imbedded bead particles directly coupling to
the transverse bundle fluctuations.

Possibilities for future studies may be to look into the effects of filament
fracture or lattice defects. The elastic energy represents a harmonic
approximation which should be extended to include nonlinear effects. Especially
for microtubules one may expected nonlinear effects to play an important role in
bundle mechanics. For example, it may be important to consider that
protofilaments in their unstressed state are not straight but bend radially
outwards. Aditional complications could also arise from the fact that some
microtubules are not transversly isotropic as we have assumed here, but have a
"seam", where protofilaments are offset relative to their neighbors. Failure
modes under axial compressive forces have been discussed in a model for
microtubules that starts from a transversly isotropic shell
theory~\cite{wang:052901}. It would be interesting to compare the results of a
generalized WLB model -- to include cross-sectional deformations -- with their
results for the critical buckling forces. One mode of failure, the ovalization
of the microtubule cross-section (Brazier effect) may for example be taken into
account by adjusting the cross-sectional coordinates with the help of an
``ovalization parameter''. This would then have to be determined together with
the other degrees of freedom from the equilibrium equations.

\acknowledgments

The authors acknowledge fruitful discussions with Andreas Bausch, Mark Bathe,
Mireille Claessens and Karen Winkler. CH acknowledges the von-Humboldt
Feodor-Lynen, the Marie-Curie Eurosim and the ANR Syscom program for financial
support. EF is grateful to the Deutsche Forschungsgemeinschaft for support
through Grant Fr 850/8-1, and to the German Excellence Initiative via the NIM
program. We also acknowledge the hospitality of the Aspen Center for Physics
where part of this work was completed.

\appendix

\section{Exact Solution for bending and twist rigidities}\label{sec:exact-solution-2d}

In this appendix the effective bending and twist rigidities are calculated exactly
from the WLB Hamiltonian, Eqs.~(\ref{eq:H0_Omega}) - (\ref{eq:Delta_mismatch}).
To this end the $u$-variables have to be integrated over to define an effective
WLC Hamiltonian.

First, we have to show that there are no terms in the shear Hamiltonian,
Eq.~(\ref{eq:Hshear}), that would couple the different Euler angles
$\theta,\phi,\psi$. To this end we use Eq.~(\ref{eq:Delta_mismatch}) in
Eq.~(\ref{eq:Hshear}) and specialize to a square cross-section. The resulting
shear Hamiltonian can then be written as
\begin{eqnarray}\label{eq:}
  H_{\rm shear} &=&\frac{k_\times}{2\delta}\int_s\sum_{ij}\left[ \left(
      u_{ij}-u_{i+1,j} -b\psi'z_j+b\phi\right)^2 \right. \nonumber\\
    &+& \left.\left(u_{ij}-u_{i,j+1} +b\psi'y_i-b\theta\right)^2\right]\,,
\end{eqnarray}
where $y_{i}= b\cdot(i + \frac{1}{2})$ and $z_{j}= b\cdot(j + \frac{1}{2})$.
Here, each filament is labeled by the pair of indices $(i,j)$, which denotes its
location in the $i$th row and the $j$th column of the square cross-section. The
only terms that couple the different Euler angles are $b^2\psi'\phi\sum_{ij}z_j$
and $b^2\psi'\theta\sum_{ij}y_i$. These identically vanish because of the
symmetric arrangement of the filaments, $\sum_j z_j = 0$ and $\sum_i y_i = 0$.

The remaining calculations are performed in Fourier-space. For the
transformation we use $\cos$-modes, which are appropriate for bundles with
pinned boundary conditions. Writing the $u$-dependent part of the Hamiltonian in
matrix form $\beta H=\frac{1}{2}\sum_{kl}u_kA_{kl}u_l+ \sum_lu_lb_l$ one needs
the following formula valid for a Gaussian integral
\begin{eqnarray}\label{eq:Hsymbolic}
  \int \prod_k du_k \exp\left(-\frac{1}{2}\sum_{kl}u_kA_{kl}u_l - \sum_lu_lb_l\right)\\\nonumber
  = \exp\left(-\frac{1}{2}\sum_{kl}\bar u_kA_{kl}\bar u_l - \sum_l\bar u_lb_l\right)\,,
\end{eqnarray}
where $\bar u$ is obtained from
\begin{equation}\label{eq:eq_for_u}
\partial (\beta H)/\partial u_k=0\,.
\end{equation}
Having found the solution $\bar u$, we can finally bring
Eq.~(\ref{eq:Hsymbolic}) in the form of Eq.~(\ref{def_eff_rigidities}) to read
off the effective bend and twist rigidities.

\subsection{Bending}\label{sec:bending}

Here we solve Eq.~(\ref{eq:eq_for_u}) for the case of bending of the square
bundle. It proves useful to introduce the dimensionless crosslink shear
stiffness $\alpha = k_\times L^2/k_s\delta^2$. We can then write
Eq.~(\ref{eq:eq_for_u}) as
\begin{equation}\label{eq:varUkFourier}
(qL)^2u_k-\alpha\partial^2u_k = 0\,.
\end{equation}
We also defined the discrete second derivative $\partial^2u_k =
u_{k+1}-2u_k+u_{k-1}$.  Note, that we are working in Fourier-space so
all quantities should carry an additional subscript relating to the
mode-number $n$. As different modes don't mix, there is no harm in
dropping it for the moment.

At the outer edges of the bundle $k=-M,M-1$ we find
\begin{equation}\label{eq:bcouter}
u_{M-1}-u_{M-2} = \frac{(qL)^2}{\alpha}u_{M-2}+b\,\theta\,,
\end{equation}
and
\begin{equation}\label{eq:bcinner}
u_{-M+1}-u_{-M}=\frac{(qL)^2}{\alpha}u_{-M}+b\,\theta\,.
\end{equation}

The Eq.~(\ref{eq:varUkFourier}) is solved by $u_k = Am_+^k+Bm_-^k$, where $m_\pm
= \frac{1}{2}\left(x\pm \sqrt{x^2-4} \right)$ and we have defined
$x=2+(qL)^2/\alpha$.  The constants $A,B$ are adjusted to fulfil the boundary
conditions Eqs.~(\ref{eq:bcouter}) and (\ref{eq:bcinner}). This solution for the
axial stretching variable $u_k$ is plotted in Fig.~\ref{fig:axialstrain}.

To obtain the approximated bending rigidity of Eq.~(\ref{eq:kappaLINEAR}) one
first has to insert the assumption $u_k=\Delta u\cdot (k+1/2)$ into the
Hamiltonian and then minimize with respect to the single variable $\Delta u$.

This way one finds 
\begin{equation}
\Delta u = \frac{-b\,\theta}{1+\frac{\sqrt{M}}{6\alpha}(\sqrt{M}+1)}\,,
\end{equation}
which has to be reinserted into the Hamiltonian to yield Eq.~(\ref{eq:kappaLINEAR}).

\subsection{Twist}\label{sec:twisting}

In the case of pure twisting the same analysis can be done to
calculate the effective twist rigidity. The difference to the bending
case is that now the full 2d boundary value problem has to be
solved. The equation determining the axial stretching $u_{ij}$ of the
filament indexed by $(i,j)$ is
\begin{eqnarray}\label{eq:equilibrium_twist}
(qL) ^{2} u_{ij}  - \alpha \Delta u_{ij} = 0,
\end{eqnarray}
where the operator $\Delta=\partial^2_i + \partial^2_j$ is the two-dimensional
version of the discrete Laplacian encountered above. The finite difference
operator $\partial_i$ is defined as $\partial^2_iu_{ij} = u_{i+1j} + u_{i-1j} -
2u_{ij} $. As for the case of pure bending, here, the twist variable $\psi$
enters only via the boundary terms.

\begin{figure}
	\centering
	\includegraphics[width=0.48\columnwidth]{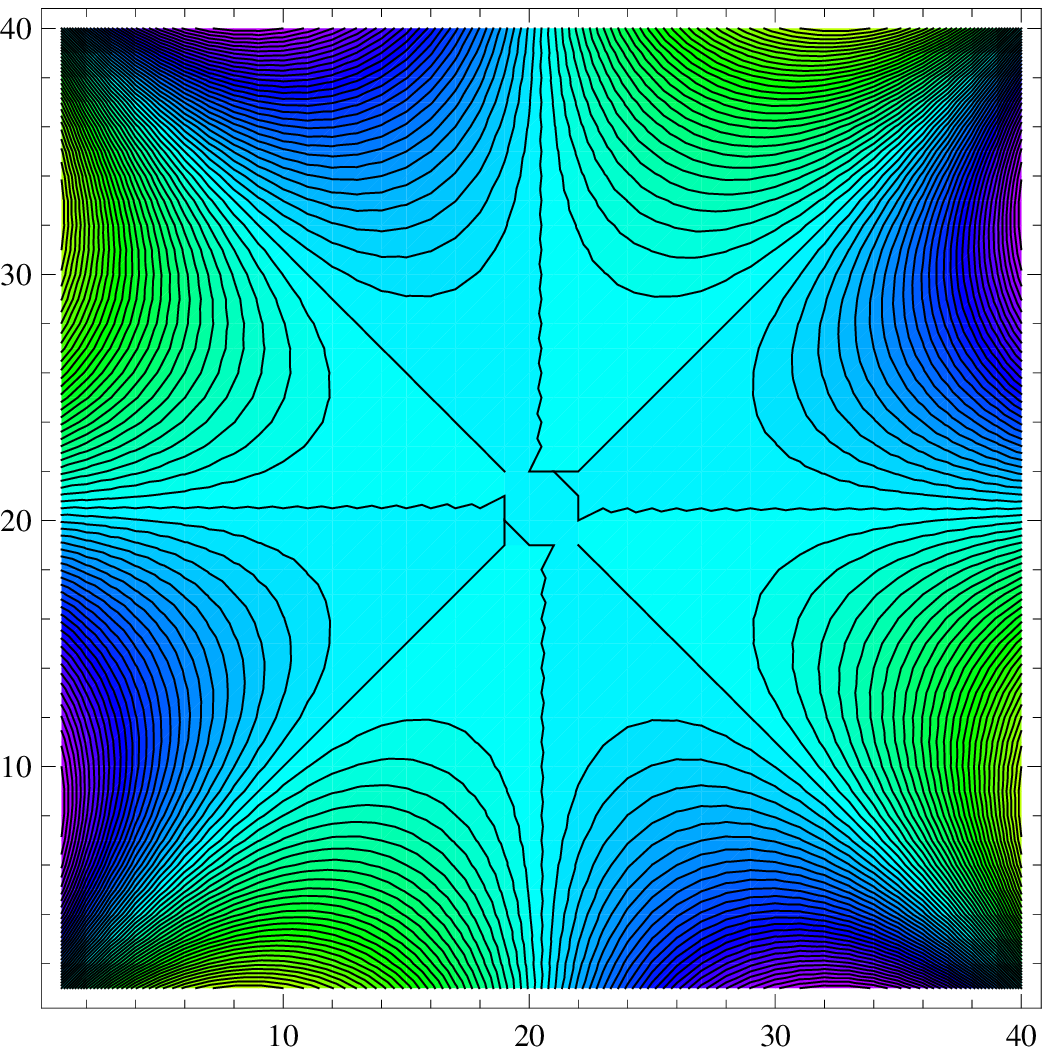}
	\hfill
	\includegraphics[width=0.48\columnwidth]{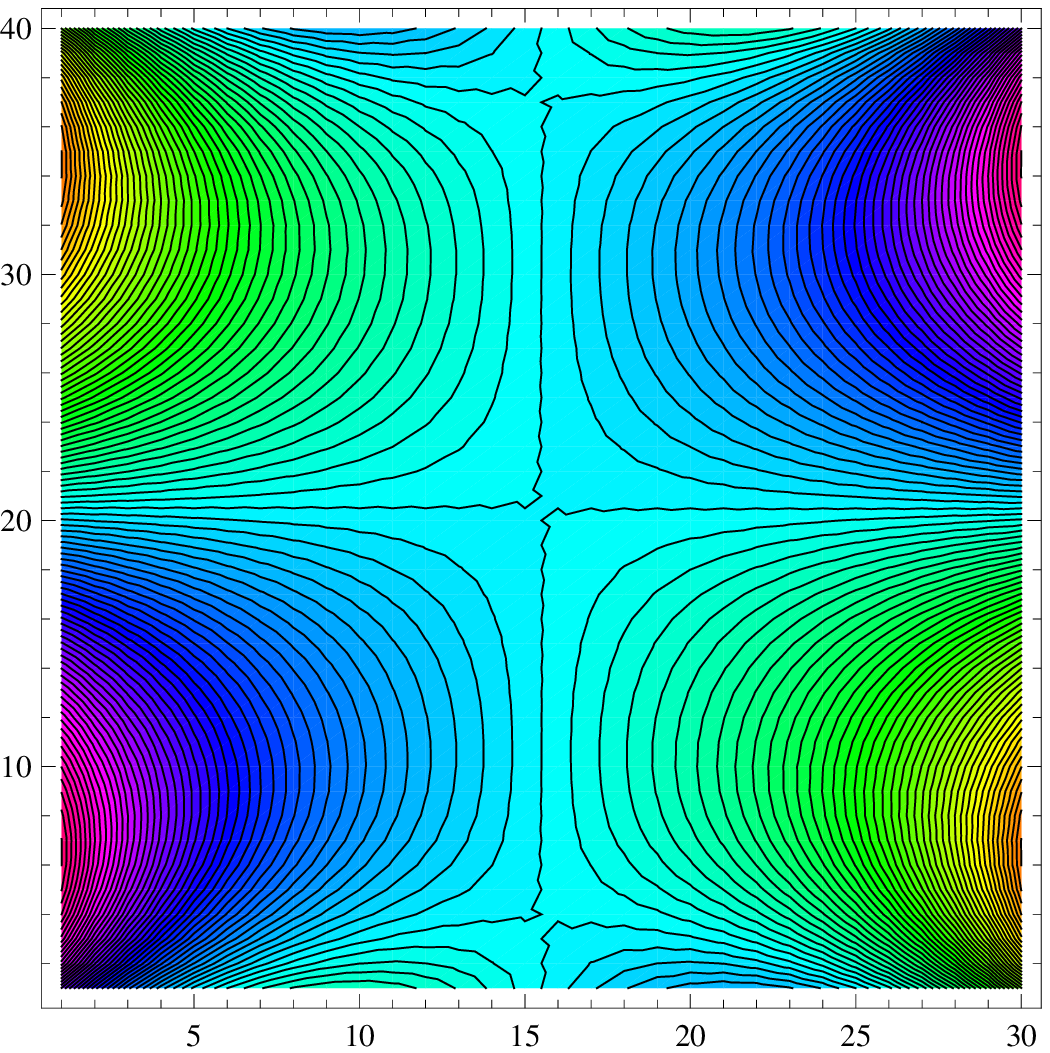}
	\caption{Illustration of the two types of solution obtained
	for the stretching $u_{ij}$ of filament $(i,j)$. Left: square
	cross-section of $40\star40$ filaments. Right: $30\star40$
	filaments } \label{fig:TorsionRechteck_20zu20}
\end{figure}

The classic theory to calculate the twist rigidity of beams has been given by
Saint-Venant~\cite{loveBOOK}. In this approach the axial displacements are found
by solving Laplace's equation $\Delta u(y,z) = 0$ on the appropriate domain of
the cross-section. We see that Eq.~(\ref{eq:equilibrium_twist}) reduces to the
Laplace equation in the limit $\alpha\sim k_\times\to\infty$, which is the
reason why in Fig.~\ref{fig:cont_limit_twist} the continuum limit is approached
with increasing shear stiffness at fixed bundle size. The remaining difference
stemming from the discretization into $N$ filaments represents only a small
effect. In Saint-Venant theory it is well known that for rectangular domains two
types of solutions appear, depending on the aspect-ratio of the cross-section.
We illustrate the different symmetry properties of these solutions in
Fig.~\ref{fig:TorsionRechteck_20zu20}.

\section{Relation between the $\Omega_{\alpha,k}$ of filament $k$ and the
  $\Omega_\alpha$ of the central line}\label{sec:relat-betw-omeg}

This appendix gives some details on the description of the bundle kinematic
degrees of freedom.  The goal is to relate the generalized curvatures
$\Omega_{\alpha,k}$ of filament $k$, to the $\Omega_\alpha$ of the bundle
central line. That these need not necessarily coincide is best illustrated by
discussing an example. If the central line is twisted but not bent, $\Omega_3$
remains as the only non-vanishing component of the curvature vector. The
filaments themselves, however, twist \emph{and} bend as they trace out a helical
path with radius $R$. Their bending energy is $\kappa_fL(R\Omega_3^2)^2$, of
fourth order in $\Omega_3$.

As a second example consider the bending of the central line, $\Omega_1= 1/R$,
where $R$ is the radius of curvature. The filaments that lie at a distance $b$
away from the central line naturally have a different radius of curvature, $R\pm
b$, and thus a different bending energy. The correction is again of higher order
and scales as $b^2\Omega_1^4$. It turns out, that all effects like the two just
mentioned only contribute to higher order. To lowest order we will find that
$\Omega_\alpha=\Omega_{\alpha,k}$.

We define the vector $\mbf{r}_k(s)=\mbf{r}_0(s)+\mbf{R}_k(s)$ to point to
filament $k$ at arclength position $s$. The central line of the bundle is
thereby given by $\mbf{r}_0(s)$. The position of each filament in the
cross-section is parametrized by $\mbf{R}_k(s) =
A_k\mbf{d}_1(s)+B_k\mbf{d}_2(s)$, where $A_k$ and $B_k$ are constants
independent of arclength $s$ and deformation of the bundle. This, in particular,
implies that in the reference state the filaments are always straight and
untwisted. We also need the derivative of $\mbf{R}_k$ with respect to $s$,
\begin{equation}\label{rkprime}
\mbf{R}_k'=(B_k\Omega_1-A_k\Omega_2)\mbf{t}+\Omega_3\mbf{R}_k^\perp
\end{equation}
where we have used the Frenet-Seret equations and defined $\mbf{R}_k^\perp =
B_k\mbf{d}_1(s)-A_k\mbf{d}_2(s)$. With this the tangent to filament $k$ is given
by
\begin{equation}\label{tangent_k} \mbf{t_k} =
\frac{1}{N}((1+B_k\Omega_1-A_k\Omega_2)\mbf{t}+\Omega_3\mbf{R}_k^\perp)\,,
\end{equation}
with an appropriate normalization factor, $N$. One finds, that the filament
tangent is parallel to the central-line only if the twist vanishes,
$\Omega_3=0$. For finite bundle twist, the filament tangent is rotated around
the vector $\mbf{R}_k$ relative to the central tangent $\mbf{t}$. The magnitude
of the rotation is $\Omega_3R_k$ and depends on the distance
$R_k=|\mbf{R}_k|=|\mbf{R}_k^\perp|$ of the filament to the central line.

In order to derive expressions for the remaining two unit vectors
$\mbf{d}_{1,k}$ and $\mbf{d}_{2,k}$ let us assume, for the time being, that no
other rotation is allowed that may re-orient the local frame of filament $k$
relative to the central frame. With this assumption filaments are not allowed to
twist relative to their neighbors as this would correspond to a rotation around
$\mbf{t}_k$. In the bundles we consider, this internal twist motion can safely
be neglected as crosslinks provide for permanent rigid inter-filament
connections. As explained below, these internal twist modes may nevertheless be
important at the time of bundle formation.

The local frame can then be related to the central frame by
\begin{eqnarray}
\mbf{d}_{1,k} &=& \mbf{d}_1-B_k\Omega_3\sin\alpha \mbf{t} + O(b_k\Omega_3)^2 \\ 
\mbf{d}_{2,k} &=& \mbf{d}_2+A_k\Omega_3\cos\alpha \mbf{t} + O(b_k\Omega_3)^2
\end{eqnarray}

One can now express the generalized curvatures of the filaments in terms of the
$\Omega_\alpha$ of the central line. For simplicity we here consider the case of pure
twisting where $\Omega_1=\Omega_2=0$. One then finds
\begin{eqnarray}
\Omega_{3,k} &\equiv& (\partial_{s_k}\mbf{d}_{1,k})\cdot\mbf{d}_{2,k} \\\nonumber
 &=& \Omega_3- \Omega_3\Omega_3'A_kB_k = \Omega_3 (1+ O(\frac{b^2\Omega_3}{L}))
\end{eqnarray}

Note, that the difference in length between the central line and the individual
filaments should be taken into account in the parametrization in terms of
arclength, however, only contributes to higher orders. More formally, $\partial
s/\partial s_k=1+O(b\Omega)$.

In addition to twist, filaments also bend
\begin{eqnarray}
  \Omega_{1,k} &\equiv& (\partial_{s_k}\mbf{d}_{2,k})\cdot\mbf{d}_{3,k} \\
    &=& \Omega_3'A_k+\Omega_3^2B_k
\end{eqnarray}
and
\begin{eqnarray}
  \Omega_{2,k} &\equiv& (\partial_{s_k}\mbf{d}_{1,k})\cdot\mbf{d}_{3,k} \\
  &=& -\Omega_3'B_k+\Omega_3^2A_k\,.
\end{eqnarray}
which reduces to the well known expression for the curvature of a helix if
$\Omega_3'=0$.

Finally, let us comment on what would happen if we did allow the filaments to
twist individually, and relative to their neighbors. This filament twist can be
described by an additional rotation, $\psi_k(s)$ of the local unit vectors,
$\mbf{d}_{1,k},\mbf{d}_{2,k}$ around the tangent $\mbf{t}_k$.

The new vectors are then given by
\begin{eqnarray}
  \mbf{e}_{1,k} &=& \cos\psi_k\mbf{d}_{1,k} +  \sin\psi_k\mbf{d}_{2,k} \\ 
  \mbf{e}_{2,k} &=&  -\sin\psi_k\mbf{d}_{1,k} +  \cos\psi_k\mbf{d}_{2,k}\,,
\end{eqnarray}
and the bundle twist, $\Omega_3$, can be calculated as before
\begin{equation}\label{eq:}
\Omega_{3,k} = \Omega_3+\psi_k'\,.
\end{equation}
The twist energy should thus be written as
\begin{eqnarray}\label{eq:}
  H_{\rm twist} &=& \frac{\kappa_t}{2}\int_s\sum_k(\Omega_3+\psi_k')^2 \\\nonumber 
  &=&  \frac{N\kappa_t}{2}\int_s(\Omega_3+\bar{\psi'})^2+N(\bar{\psi'^2}-{\bar{\psi'}}^2) \,,
\end{eqnarray}
where we defined the moments of the filament twist distribution $\bar{\psi'^r}=
\sum_k{\psi_k'}^r/N$.

As explained above, in crosslinked bundles this internal twist can be assumed
to be quenched at the time of bundle formation. In this case we need not treat
the $\psi_k$ as dynamical variables for the discussion of bundle deformation. A
finite $\bar{\psi'}$ nevertheless gives the bundle a certain helicity and
imposes pretwist, as discussed in Section~\ref{sec:pretwisted-bundles}.

\section{Derivation of shear deformation}\label{sec:deriv-shear-deform}

To calculate the geometric part $\Delta$ of the shear deformation, an expression
for the arclength mismatch between the two points on the filament pair is
needed. The general expression is
\begin{equation}
  \Delta_{lk} = \Delta s + \int^s|\mbf{t}+\mbf{R}_l'| - \int^s|\mbf{t}+\mbf{R}_k'|\,,
\end{equation}
where the first contribution, $\Delta s$, derives from the possibility that the
two points do not correspond to the same point, $s$, on the bundle central line.

By using Eqs.~(\ref{rkprime}), (\ref{tangent_k}) and expanding one finds
\begin{eqnarray} 
\Delta = \Delta s + \int_0^s\left[\mbf{b}_{lk}'\cdot
  \mbf{t}-\frac{1}{2}(\mbf{R}_{l}'^2-\mbf{R}_{k}'^2)\right]\,,
\end{eqnarray}
where, we defined $\mbf{b}_{lk}=\mbf{R}_{l}-\mbf{R}_{k}$ pointing from filament
$k$ to filament $l$. As in the formulation of $H_0$, only the leading order
terms have been accounted for. The $(\mbf{b}_{lk}'\cdot \mbf{t})$-term only
contains bending deformations. This may be seen by setting $\mbf{t}=\mbf{\hat
  e}_x=\rm{const}$ appropriate for a pure twisting of the central line. Then
$\Delta_{lk} = \mbf{b}_{lk}\cdot \mbf{\hat e}_x=0$ as the vector $\mbf{b}_{lk}$
lies within the cross-section, that is perpendicular to the tangent.

The last term corresponds to the arclength-difference acquired between two
filaments at different distance to the central-line ($R_k\neq R_l$). It is clear
that the filament farther out has to go a longer distance, so its crosslinking
sites will stay back in comparison to that of its neighbor closer in the center
of the bundle. However, twist also produces shear deformation between filaments
which lie at equal distance to the central line ($R_l=R_k$). This is embodied in
the extra term $\Delta s$, which we treat now (also see
Fig.~\ref{fig:delta_mt}). To derive an expression for the shear displacement in
this case, assume that the two filaments lie, separated by a distance
$b_{lk}=|\mbf{b}_{lk}|$, on the surface of a cylinder of radius $R$. Twisting
the cylinder makes the filaments wind around it, each taking an angle
$\Omega_3R$ to the cylinder axis.  The shear deformation then simply is $\Delta
s = \Omega_3 R b_{lk}$. For arbitrary orientation of the filament pair we have
to write $\Delta s = \mbf{t}_{lk}\cdot\mbf b_{lk}$, where $\mbf{t}_{lk}$ is the
tangent to the midline between the filament pair. In agreement to what has been
said above, this mechanism does not contribute to the shear displacement when
the filament pair $(l,k)$ is connected by crosslinks in radial direction. In
this case, $\mbf{t}_{lk} \perp \mbf{b}_{lk}$, and the shear deformation $\Delta
s=0$.


\end{document}